\documentclass{article}[12pt]

\usepackage{amsmath, amssymb, amsfonts, mathrsfs, bbm}
\usepackage[T1]{fontenc}

\usepackage[final]{graphicx}
\usepackage{color}
\usepackage{array}
\usepackage{colortbl}
\usepackage{subfigure}
\usepackage{hep}
\usepackage{float}

\def\N{{\cal N}}
\def\pr{{\partial}}
\def\tr{{\mathrm{tr}}}

\begin{document}

\thispagestyle{empty}

\begin{center}
{\bf \Large

$\phantom{+}$ \vspace{1cm}

 Introduction to Gauge/Gravity Duality

\medskip

(TASI lectures 2017)

}
\end{center}
%\ShortTitle{Introduction to Gauge/Gravity Duality}

\bigskip

\bigskip 

\bigskip

\begin{center} {\bf \large
{Johanna Erdmenger}}
        
\bigskip

Institut f\"ur Theoretische Physik und Astrophysik \\
       Julius-Maximilians-Universit\"at  W\"urzburg 

\bigskip 

\bigskip

\bigskip

\end{center}

We review how the AdS/CFT correspondence is motivated within
string theory, and discuss how it is generalized to gauge/gravity
duality. In particular, we highlight the relation to quantum
information theory by pointing out that the Fisher information metric
of a Gaussian probability distribution corresponds to an Anti-de
Sitter space.  As an application example of gauge/gravity duality, we present a holographic Kondo
model. The Kondo model in condensed matter physics describes a spin
impurity interacting with a free electron gas: At low energies, the
impurity is screened and there is a logarithmic rise of the
resistivity. In quantum field theory, this amounts to a negative beta
function for the impurity coupling and the theory flows to a
non-trivial IR fixed point. For constructing a gravity dual, we
consider a large $N$ version of this model in which the ambient
electrons are strongly coupled even before the interaction with the
impurity is switched on.  We present the brane construction which
motivates a gravity dual Kondo model and use this model to calculate the
impurity entanglement entropy and the resistivity, which has a
power-law behaviour. We also study quantum quenches, and discuss the
relation to the Sachdev-Ye-Kitaev model. 

%{Lectures given at the TASI School 2017. }

\bigskip

Lectures given at the Theoretical Advanced Study Institute (TASI) Summer School 2017 "Physics at the Fundamental Frontier",
         4 June - 1 July 2017, 
         Boulder, Colorado.

\newpage

\tableofcontents

\section{Introduction}

Gauge/gravity duality is one of the major developments in theoretical
physics over the last two decades. Based on string theory, it provides
a new relation between a quantum field theory without gravity and a
gravity theory itself. At a fundamental level within theoretical
physics, this has provided new insight into the nature
of quantum gravity. Moreover, since in a certain limit gauge/gravity
duality maps strongly coupled quantum field theories
to classical gravity theories, it 
has provided a new way for calculating observables in these strongly
coupled theories which are generically hard to solve. Gauge/gravity
thus provides new unexpected links between previously unrelated areas
of physics.

\begin{figure}[h]
\centering
{\includegraphics[scale=0.3]{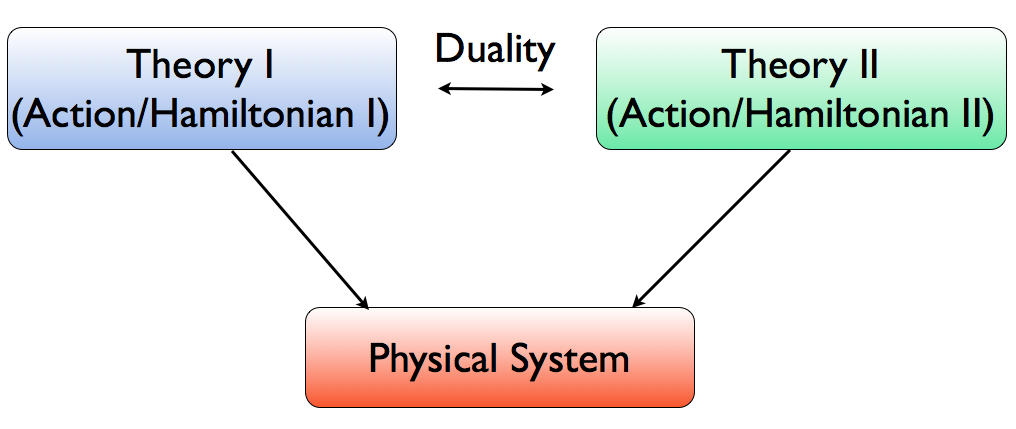} }
\caption{Two physical theories describing the same physical system are
related by a {\it duality}. }
\label{figure1}
\end{figure}

What is a duality? Imagine that a physical system is described by two
different actions or Hamiltonians that may involve different encodings
of the degrees of freedom. Then, these two different theories are said
to be related by a {\it duality}. This is visualized in figure
\ref{figure1}.
There are many well-known examples for dualities in physics. One of
these is the duality between the massive Thirring model and the
sine-Gordon model within two-dimensional quantum field theory
\cite{1975PhRvD..11.2088C}. This is a boson-fermion duality within
quantum field theory.  A
further example is Montonen-Olive duality of electric and magnetic
charges \cite{1977PhLB...72..117M}, which is an example of a duality
between a weakly and a strongly coupled gauge theory or {\it
  S-duality}.   This plays also an important role in string theory,
together with T-duality \cite{1985NuPhB.258...46C}.

{\it Gauge/gravity duality}, as first realized by the AdS/CFT
correspondence of Maldacena \cite{Maldacena:1997re},  is a very special duality in the sense
that it relates a gravity theory to a gauge theory, i.e.~a quantum field theory without
gravity. This new relation implies new questions about the nature of
gravity itself: How is gravity related to quantum physics? It is
equivalent to a non-gravity theory at least in this special context -
does this imply that it is non-fundamental? This is an open question
which we will not explore in detail here. Nevertheless we note that
gauge/gravity duality opens up new issues about the nature of
gravity. It is important to emphasize in this context that so far the
best understood examples of gauge/gravity duality involve gravity
theories with negative cosmological constant, different from the
theory describing our Universe in which the cosmological constant is
extremely small but positive.

A further important aspect is that generically, gauge/gravity duality
relates a quantum gauge theory in flat space to a string theory in
curved space. Only in a very particular limit, which we will discuss
in detail, this string theory reduces to a classical gravity theory of
pointlike particles. On the field theory side, this same limit implies
that the quantum gauge theory becomes strongly coupled and the rank of
its gauge group goes to infinity.

{\it Applications of gauge/gravity duality.} The fact that
gauge/gravity duality relates strongly coupled quantum field theories
to weakly coupled classical gravity theories provides a new approach
to calculating observables in these strongly coupled quantum field
theories. Generically, such theories are hard to study since there is
no universal approach for calculating observables in them. This is
crucially different from weakly coupled quantum field theories, for which
perturbation theory is the method of choice and provides very accurate
results. An example for an approach to strongly coupled gauge theories
is lattice gauge theory, in which space-time is discretized and
advanced numerical methods are used. Lattice gauge theory is very
successful in calculating observables such as bound state masses,
however it is afflicted by the sign problem which renders the
description of transport properties very complicated, in particular at
finite temperature and density. It is thus desirable to have an
alternative approach at hand which allows for
comparison. Gauge/gravity duality provides such an approach. 

Strongly coupled quantum field theories appear in all areas of
physics, including condensed matter physics. Weakly coupled
theories may successfully be described in a quasiparticle
approach. Quasiparticles are quantum excitationd in one-to-one
correspondence with the states in the corresponding free
(non-interacting) theory. In strongly-coupled systems however, this
map is no longer present. In general, the excitations in these systems
are collective modes of the individual degrees of
freedom. Gauge/gravity duality provides an elegant way of describing
these modes by mapping them to {\it quasinormal modes} of the gravity
theory. These modes are complex eigenfrequencies of the fluctuations
about the gravity background: Their real part is related to the mass
of the fluctuations and their complex part to the decay width.

Before we proceed, it is important to stress that to the present day,
gauge/gravity duality is a conjecture which has not been proved. The
proof is hard in particular since it would require a non-perturbative
understanding of string theory in a curved space background, which is
not available so far.

These lecture notes only give an outline of the most important concepts. 
Detailed information on gauge/gravity duality, the AdS/CFT
correspondence and its applications may be found for instance in the
books
\cite{Ammon:2015wua,Nastase:2015wjb,Natsuume:2014sfa,Mateos,Zaanen:2015oix}. 
There are also very useful lecture notes of previous TASI schools, see
for instance \cite{Klebanov:2000me,Maldacena:2003nj}.  Further lecture
notes on AdS/CFT include  \cite{Ramallo:2013bua,Petersen:1999zh,Aharony:1999ti}.

 In the present notes, we also
include comments on recent developments relating the AdS/CFT
correspondence to concepts from quantum information.
In the second part of these lectures, we focus on the Kondo model and
a variant of it with a gravity dual. This provides a new example for
constructing a gravity dual, and its applications. This provides a
further entry in the list of examples of gauge/gravity
duality. Related further lectures at TASI 2017 are those by Harlow
\cite{Harlow:2018fse} and DeWolfe \cite{DeWolfe:2018dkl}  in particular.

\section{AdS/CFT correspondence}

\setcounter{equation}{0}

\subsection{Statement of the correspondence}

Let us begin by considering the best understood example of
gauge/gravity duality, the AdS/CFT correspondence. Here, `AdS' stands
for `Anti-de Sitter space' and and `CFT' for `conformal field
theory. The Dutch physicist Willem de Sitter was a  friend of
Einstein. The prefix `Anti' refers to the fact that a crucial sign
changes from plus to minus. In fact, Anti-de Sitter space is a
hyperbolic space with a negative cosmological constant.

In this example  a
four-dimensional CFT, $\N=4$ $SU(N)$  Super
Yang-Mills theory,  is conjectured to be dual to gravity in the space AdS$_5 \times$
S$^5$. This was proposed along with other examples for AdS/CFT by
Maldacena in his seminal paper \cite{Maldacena:1997re} in 1997.  As we will see, the two theories have the same amount of degrees of
freedom per unit volume and the same global symmetries.
 We will first state the duality and then explain it
in detail. The AdS/CFT correspondence states that

\bigskip

\parbox{0.95\linewidth}{
{\bf $\mathcal{N}=4$ Super Yang-Mills (SYM) theory}  with gauge group $SU(N)$ and Yang-Mills coupling  $g_\mathrm{YM}$\\[2mm]
{\bf is dynamically equivalent to}\\[2mm]
{\bf Type IIB superstring theory on AdS$_5 \times S^5$}, with string length $l_s =
\sqrt{\alpha^\prime}$ and coupling constant $g_s$.  The radius of
curvature of both AdS$_5$ and $S^5$ is $L$, and there are $N$ units of $F_{(5)}$ flux on $S^5.$\\[2mm]
The two free parameters on the field-theory side, i.e.~$g_\mathrm{YM}$
and $N$, are related to the free parameters $g_s$ and $L / \sqrt{\alpha^\prime}$ on the string theory side by
\begin{equation}
g_\mathrm{YM}^2 = 2 \pi g_s  \quad \mbox{and} \quad 2
g_\mathrm{YM}^2 N  = L^4/\alpha^{\prime  2}. \nonumber
\end{equation}
}

\noindent For understanding this duality and its motivation in detail, let us
first recall some properties of the ingredients involved. We begin
with the field theory side and introduce conformal field theories and
$\N=4$ supersymmetry.

\subsection{Prerequisites for AdS/CFT}

\subsubsection{Conformal symmetry}

An essential aspect for the AdS/CFT correspondence is that the quantum field
theory involved is a conformal field theory (CFT). Such a theory
consists of fields that transform covariantly under conformal
coordinate transformation. These leave angles invariant (locally) and
in flat  $d$-dimensional spacetime 
are defined by the following transformation law of the metric,
\begin{equation}
dx'_\mu dx'^\mu = \Omega^{-2} (x) dx_\mu dx^\mu \, .
\end{equation}
Infinitesimally, with  $\Omega(x) = 1 -\sigma (x)$ and $x'^\mu = x^\mu
+ v^\mu (x)$, this gives rise to the conformal Killing equation
\begin{equation} \label{cK}
\pr_\mu v_\nu + \pr_\nu v_\mu = 2 \sigma(x) \eta_{\mu \nu} \, , \qquad
\sigma (x) = \frac{1}{d} \pr \cdot v \, .
\end{equation}
In $d=2$ dimensions, this reduces to the Cauchy-Riemann equations,
which are solved by any holomorphic function. This implies that in
$d=2$, conformal symmetry is infinite dimensional and thus leads to an
infinite number of conserved quantities. In more than two dimensions
however, conformal symmetry is finite dimensional and the only
solutions to the conformal Killing equation \eqref{cK} are
\begin{equation} \label{conftr}
v^\mu (x) = a^\mu + \omega^{\mu}{}_{ \nu} x^\nu + \lambda x^\mu + b^\mu x^2
- 2 (b \cdot x) x^\mu \, ; \quad \omega_{\mu \nu} = - \omega_{\nu \mu}
\, , \sigma(x) = \lambda - 2 b\cdot x\,  .
\end{equation}
In $d>2$, the conformal Killing vector $v_\mu (x)$ is at most
quadratic in $x$. It contains translations (of zeroth order in $x$),
rotations and scale transformations (both linear in $x$) and special
conformal transformations (quadratic in $x$). The scalar $\lambda$,
the vectors $a_\mu$ and $b_\mu$ and the antisymmetric matrix
$\omega_{\mu \nu}$ contain a total of
\begin{equation}
1 + 2d + d (d-1)/2 = (d+1) (d+2)/2
\end{equation}
free parameters. In Euclidean signature, the symmetry group generated
by these transformations is $SO(d+1,1)$, while in Lorentzian
signature, it is $SO(d,2)$. Let us examine the algebra
associated to
the infinitesimal transformations \eqref{conftr}  with parameters
$(a^\mu, \omega^{\mu \nu}, \lambda, b^\mu)$ for
the Lorentzian case. The generator for translations  is the momentum
operator $P_\mu$. The generator for Lorentz transformations is denoted 
by $L_{\mu\nu}$. The generator for scale transformations is $D$ and the
generator for special conformal transformations 
is $K_\mu$. The
conformal algebra consists of the Poincar\'e algebra supplemented by
the relations
\begin{align} 
[L_{\mu \nu}, K_\rho ] & = i (\eta_{\mu \rho} K_\nu - \eta_{\nu \rho}
  K_\mu ) \, , \qquad [ D, P_\mu ] = i P_\mu \, ,  \\
[D, K_\mu]  & = -i K_\mu\, , \qquad [D, L_{\mu \nu}] = 0 \, , \qquad
  [K_\mu, K_\nu] = 0 \, , 
\\ [ K_\mu , P_\nu] &= -2 i (\eta_{\mu \nu} D -
  L_{\mu \nu}) \, .
\end{align}
For the representations we postulate
\begin{equation}
[ D, \phi (0) ] = -i \Delta \phi(0)
\end{equation}
for any field $\phi(x)$. This implies
\begin{equation}
\phi(x) \rightarrow \phi' (x') = \lambda^{-\Delta} \phi (x)
\end{equation}
for $x \rightarrow x' = \lambda x$. $\Delta$ is the scaling dimension
of the field $\phi$. For an infinitesimal transformation this gives
\begin{equation}
\delta_D \phi \equiv [ D, \phi(x) ] =-i \Delta \phi(x) - i x^\mu
\pr_\mu \phi (x) \, ,
\end{equation}
with similar relation for the other conformal transformations
$\delta_P \phi$, $\delta_L \phi$, $\delta_K \phi$.

For organising the representations, it is useful to define the {\it
quasiprimary} fields which satisfy
\begin{equation}
[ K_\mu, \phi(0) ] = 0 \, .
\end{equation}
This defines the fields of lowest scaling dimension in an irreducible representation
 of the conformal algebra. All other fields in this
multiplet, the conformal {\it descendents} of  $\phi$, are obtained by
acting with $P_\mu$ on the quasiprimary fields.

The infinitesimal transformations $\delta \phi$ give rise to the {\it
  conformal Ward identities} 
\begin{equation}
\sum\limits_{i=1}^n \langle \phi_1 (x_1) \dots \delta \phi_i (x_i)
\dots \phi_n (x_n)  \rangle = 0 \, .
\end{equation}
For scalar conformal fields this implies
\begin{gather}
\langle \phi_1 (x_1) \phi_2 (x_2) \rangle =  \begin{cases}
                                                      \frac{c}{(x_1
                                                      -x_2)^{2 \Delta}}
                                                      & \text{if}
                                                        \;\; \Delta_1
                                                        = \Delta_2 =
                                                        \Delta \, , \\
                                                      0 &
                                                          \text{otherwise.} \end{cases} 
\end{gather}
For fields with spin, the conformal transformation acts also on the
spacetime indices and reads
\begin{equation}
\delta_v \mathcal{O} (x) = - {\cal L}_v  \mathcal{O} (x) \, , \qquad {\cal L}_v \equiv
v \cdot \pr_x  + \frac{\Delta}{d} \pr \cdot v - \frac{i}{2}
\pr^{[\mu} v^{\nu]}   L_{\mu \nu} \, ,
\end{equation}
for an operator $\mathcal{O}(x) $ of arbitrary spin. The Lorentz
generator $L_{\mu \nu}$ acts on the spin indices.
For these operators, the conformal correlation functions are more
involved. However, conformal symmetry still fixes the two- and three
point functions up to a small
number of independent contributions \cite{Osborn:1993cr,Erdmenger:1996yc}.

\subsubsection{$\N=4$ Supersymmetry}

The $\N=4$ $SU(N)$ Super-Yang-Mills theory has some very special
properties which are at the origin of it possessing a gravity
dual. First of all, it was shown \cite{Howe:1983wj,Howe:1983sr} that this theory is
conformally invariant even when quantised; its beta function vanishes
to all orders in perturbation theory and also non-perturbative
contributions are absent. A further important property is that this
theory has a global $SU(4)$ symmetry, which is isomorphic to
$SO(6)$. We will see that both the $SO(4,2)$ conformal symmetry as
well as $SU(4)$ are also realized as isometries in the dual gravity
theory.  We also note that $\N=4$ Super Yang-Mills theory is invariant
under S-duality \cite{Osborn:1979tq}. 

For the $\N=4$ theory, the global $SU(4)$ symmetry is realized as an R
symmetry of the supersymmetry algebra. This algebra has four
supersymmetry generators which satisfy the anticommutation relations
\begin{gather} \label{Qcom}
\{ Q^a_\alpha , \bar Q_{b \dot \beta } \} = 2 \sigma^\mu{}_{\alpha \dot \beta}
P_\mu \delta^a{}_b \, , \qquad a = 1,2,3,4  \, ,
\end{gather}
with $\sigma^\mu{} = (\mathbbm{1}, \vec{\sigma})$ and $\vec{\sigma}$
the three Pauli matrices.
\eqref{Qcom} is invariant under  $SU(4)$ rotations. This algebra may be
combined with the conformal algebra into a superconformal
algebra. This requires the introduction of further fermionic generators, the
special superconformal generators $S^a_\alpha$ that satisfies
\begin{gather} \label{Scom}
\{ S^a_\alpha , \bar S_{b \dot \beta} \} = 2 \sigma^\mu{}_{\alpha \dot \beta}
K_\mu \delta^a{}_b \, , \qquad a = 1,2,3,4 , 
\end{gather}
with $K_\mu$ the generator of special conformal transformations. We
note that the anticommutation relation for the generators $S^a_\alpha$
\eqref{Scom} is formally similar to the one for the generators
$Q^a_\alpha$ given by \eqref{Qcom}, with the momentum operator $P_\mu$
replaced by the special conformal transformations $K_\mu$. The
operators $P_\mu, L_{\mu \nu}, D, K_\mu$ together with the
$Q^a_\alpha, S^a_\alpha$ form the superconformal algebra associated to
the superconformal group $SU(2,2|4). $

\begin{table}
\begin{center}
\begin{tabular}{l|c|c}
{Fields} & & $SU(4)$ {rep.} \\ \hline 
{Gauge field} & $A_\mu$ & $\mathbf{1}$ \\
{Complex fermions} & $\lambda^a_\alpha$  & $\mathbf{4} $ \\
{Real scalars} &$ X^i$ & $\mathbf{6} $\\ \hline
\end{tabular}
\caption{Supermultiplet of \N=4 Supersymmetry.}
\label{super}
\end{center}
\end{table}

%%%%% TABLE N=4 multiplet

The elementary fields of $\N=4$ Super Yang-Mills theory are organized
in a single multiplet of $SU(4)$, as shown in table \ref{super}. The $SU(N)$ gauge field is a singlet of
$SU(4)$. Moreover, the supermultiplet involves four complex Weyl fermions
$\lambda^a_\alpha$ in the fundamental representation ${\bf 4}$ of
$SU(4)$ and six real scalars $X^i$ in the representation ${\bf 6}$ of
$SU(4)$. Note that due to the supersymmetry, both the Weyl fermions
and the scalars are in the adjoint representation of the gauge group
$SU(N)$ since they are in the same multiplet as the gauge field.

The action of $\N=4$ Super Yang-Mills theory reads
\begin{align}  \label{N4action}
S & = \mathrm{tr} \int \! d^4 x \left( - \frac{1}{2g_\mathrm{YM}{} ^2} F_{\mu \nu}
    F^{\mu \nu} - i \sum\limits_{a=1}^{4}\bar \lambda^a \bar \sigma^\mu D_\mu \lambda_a - 
\sum\limits_{i=1}^6 D_\mu \phi^i D^\mu \phi^i  \right. \nonumber\\ &
                                                                     \; \hspace{5mm}
                                                                     \left. +
                                                                     g_\mathrm{YM}{} \sum\limits_{a,b,i} 
    C^{ab}{}_i \lambda_a [ \phi^i, \lambda_b] + g_\mathrm{YM}{} \sum\limits_{a,b,i}
    \bar C_{iab} \bar \lambda^a [ \phi^i, \bar \lambda^b]  +
    \frac{g_\mathrm{YM}{}^2}{2} \sum\limits_{i,j} [ \phi^i, \phi^j]^2 \right) \, ,
\end{align}
with $gg_\mathrm{YM}$ the Yang-Mills coupling.
The $C^{ab}_i $ are Clebsch-Gordan coefficients that couple two $\bf
4$ representations to one $\bf 6$ representation of the algebtra of 
$SU(4)_R$. We note that in addition to the kinetic terms, this action
contains interactions between three and four gauge fields via the
non-abelian gauge-field commutators in $F^{\mu \nu}$, as well as
Yukawa interaction terms between two fermions and a scalar, and a
quartic scalar interaction.

\subsubsection{Large $N$ limit}

The large $N$ limit plays an essential role for the AdS/CFT correspondence. It corresponds to a saddle point approximation. As realized by 't Hooft in 1974 \cite{Hooft74}, the perturbative expansion of fields in the adjoint representation of the $SU(N)$ gauge group may be reorganized  using a double-line notation.

\begin{figure}
\begin{center}
\includegraphics[scale=0.3]{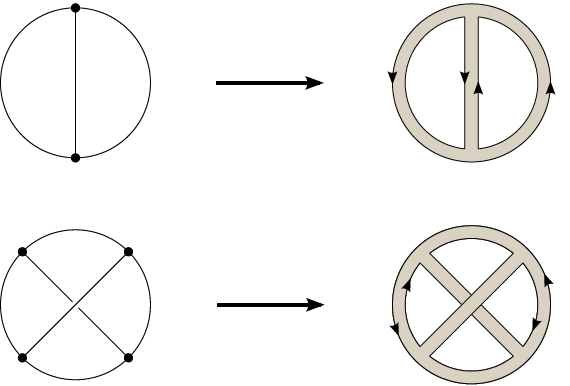}
\caption{Replacing single-line by double-line Feynman diagrams. Top:
  Planar diagram of genus zero. Bottom: Non-planar diagram of genus
  one.}
\end{center}
\label{figN}
\end{figure}

A field $\phi$ in the adjoint representation may be written as
\begin{equation}
\phi = \phi^A T^A \; \Leftrightarrow \;  (\phi)^i{}_j = \phi^A (T^A)^i{}_j \, ,
\end{equation}
where the $T^A$ are the $N^2-1$ generators of $SU(N)$. These are matrices with indices $i,j$.
If $\phi$ is a scalar field in 3+1 dimensions, then its propagator in configuration space is given by
\begin{equation} \label{prop}
\langle \phi^i{}_j (x) \phi^k{}_l (y) \rangle \, = \, \delta^i{}_l \delta^k{}_j \frac{ g^2}{ 4 \pi^2 (x-y)^2} \, ,
\end{equation}
where $g$ is a typical coupling in the theory. The Kronecker deltas
enter from the $SU(N)$ completeness relation 
\begin{equation} \label{complete}
\sum\limits_{A=1}^{N^2-1} (T^A)^i{}_j (T^A)^k{}_l =\delta^i{}_l
\delta^k{}_j - \frac{1}{N} \delta^i{}_j
\delta^k{}_l \, ,
\end{equation}
in which the second term is suppressed for $N \rightarrow \infty$. 
%%%%%%%% Normalization of vertices $1/g^2$ 
For scalar fields, $g$ in \eqref{prop}
may be the coupling of a cubic interaction term; a quartic interaction
term may then enter with coefficient $g^2$.
 In Yang-Mills theory,
$g$ will be the gauge coupling. It will turn out to be extremely useful to 
define the {\it 't~Hooft coupling} 
\begin{equation}
\lambda = g^2 N \, .
\end{equation}
Let us now count how the contributions corresponding to Feynman
diagrams scale with $N$ and with $\lambda$. Note that in the
normalization for the propagators chosen in \eqref{prop}, the vertices
scale as $1/g^2$. Also, the sum over traces of indices contributes a
factor of $N$ for every closed loop. Assembling all the ingredients,
we find that the Feynman diagrams scale as
\begin{equation}
f (\lambda, N) \sim N^{V-E+F} \lambda^{E-V}  = N^\chi \lambda^{E-V} \, ,
\end{equation}
where $V$, $E$ and $F$ are the numbers of vertices, edges and faces of
the surfaces created by the Feynman diagrams, respectively. $\chi$ is
the Euler characteristic given by
\begin{equation}
\chi = V-E+F = 2 - 2 G \, ,
\end{equation}
with $G$ the genus of the surface. We see that the leading order in
$N$ is given by $G=0$, i.e.~by the planar diagrams. Two examples for
double-line vacuum Feynman diagrams are given in figure \ref{figN}: 
The planar diagram scales as $N^2$ while the non-planar diagrams is
$1/N^2$ suppressed and scales as $N^0$. 
We note that the Feynman diagrams shown look like string-theory
diagrams with strings splitting and joining. This provides a hint that
large $N$ quantum field theories are related to string theories. In
the simple example with scalar fields considered here, it is not
possible to determine exactly which string theory is given by the collection
of large $N$ field-theory Feynman diagrams. The AdS/CFT
correspondence however provides a map between well-defined field
theories and string theories.

%%%%%% Figures planar/ non planar vacuum diagrams

\subsubsection{AdS spaces}

Anti-de Sitter (AdS) spaces play an important role in the AdS/CFT
correspondence. This has several reasons: First of all, the isometries of
AdS space in $d+1$ dimensions form the group $SO(d,2)$, which
corresponds to the conformal group of a CFT in $d$
dimensions. Moreover, AdS space has a constant negative curvature and
a boundary at which we may imagine this CFT to be defined. 

The embedding of $(d+1)$-dimensional AdS space into
$(d+2)$-dimensional flat Minkowski spacetime is provided by the
surface satisfying
\begin{gather} \label{hyperboloid}
X_1{} ^2 + X_2{}^2 + \dots + X_d{}^2 - X_0{}^2 - X_{d+1}{}^2 =  - L^2 \, ,
\end{gather}
where $X_0$, $X_1$, $\dots$ $X_{d+1}$ are the coordinates of
$(d+2)$-dimensional Minkowski space. $L$ is referred to as the {\it AdS}
{\it radius}. We note that in Lorentzian
signature, the symmetry of the isometries of AdS$_{d+1}$ is thus
$SO(d,2)$, which coincides with the symmetry of a CFT$_d$, i.e.~a
conformal field theory in $d$ dimensions with Lorentzian signature. In Euclidean signature, the
sign in front of $X_0{}^2$ becomes a plus and the symmetry is
$SO(d+1,1)$. 

The boundary of AdS$_{d+1}$ is located at the limit of all coordinates
$X_M$ becoming asympto\-ti\-cally large.
For large $X_M$, the hyperboloid given by \eqref{hyperboloid}
approaches the light-cone in $\mathbbm{R}^{d,2}$, given by
\begin{gather} \label{h2}
-X_0{}^2 + \sum\limits_{i=1}^{d} X_i{}^2 - X_{d+1}{}^2 = 0 \, .
\end{gather}
The boundary corresponds to the set of all lines on the light cone
given by \eqref{h2} which originate from the origin of
$\mathbbm{R}^{d,2}$, i.e.~$0 \in \mathbbm{R}^{d,2}$. This space
corresponds to a conformal compactification of Minkowski space.

A set of coordinates that solves \eqref{hyperboloid}
is
\begin{equation}\label{eq:Global}
\begin{aligned}
X^0 \ &= \ L \, \cosh \rho \, \cos \tau \, ,\\
X^{d+1} \ &=  \ L \, \cosh \rho \, \sin \tau \, ,\\
X^i \ &= \ L\, \Omega_i \, \sinh \rho \, , \qquad \mbox{for} \;
i=1,\dots, d \, ,
\end{aligned}
\end{equation}
where $\Omega_i$ with $i=1,\ldots, d$ are angular coordinates satisfying
$\sum_i\Omega_i^2 = 1.$The remaining coordinates take the ranges $\rho
\in \mathbbm{R}_+$ and $\tau \in [0,2\pi[.$ The coordinates $(\rho,
\tau, \Omega_i)$ are referred to as {\it global coordinates} of
AdS$_{d+1}$. It is convenient to introduce a new coordinate $\theta$ by
$\tan \theta = \sinh \rho.$ 
Then the metric associated to the parametrization (\ref{eq:Global}) becomes that of the Einstein static universe $\mathbbm{R} \times S^d$,
\begin{equation} \label{eq:theta}
d s^2 =  \frac{L^2}{\cos^2 \theta}  \; \big( - \, d\tau^2 \ + \ d\theta^2 \ + \ \sin^2 \theta\, d\Omega_{d-1}^2 \big) \, .
\end{equation}
Since $0 \leq \theta < \frac{\pi}{2}$, this metric covers
 half of $\mathbbm{R} \times S^{d}$.

\begin{figure}
\begin{center}
\includegraphics[scale=0.45]{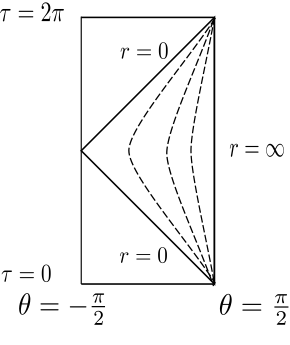}
\caption{Within AdS$_2$, the Poincar\'e coordinates cover the triangular region
  shown. The dashed lines correspond to fixed constant values of
  $r$. The boundary is at $r=\infty$. $\theta$ and $\tau$ are as defined
  in \eqref{eq:theta}; for AdS$_2$, the coordinate $\theta$ ranges
  from $- \pi/2 \leq \theta \leq \pi/2$ since the sphere $S^0$
  consists of two points. }
\label{adsfig}
\end{center}
\end{figure}

It is often useful to consider a metric in local coordinates on
AdS$_{d+1}$. This is obtained from the parametrization, with $\vec{x}
= (x^1, \dots , x^{d-1})$, 
\begin{align}
X_0 &= \frac{L^2}{2r} \left(1 + \frac{r^2}{L^4} (\vec{x}^2 - t^2 + L^2)
  \right) \, , \nonumber\\ 
X_i &= \frac{ rx_i}{L}  \;  \; \mathrm{for} \; \; i \in \{ 1, \dots,
  d-1\} \, , \nonumber\\
X_d & = \frac{L^2}{2r} \left(1 + \frac{r^2}{L^4} (\vec{x}^2 - t^2 -
      L^2) \right)\, , \nonumber\\ 
X_{d+1} & = \frac{rt}{L} \, .
\end{align}
This covers only one half of the AdS spacetime since $r > 0$.   The
corresponding metric is referred to as {\it Poincar\'e  metric} and
reads
\begin{gather} \label{poin}
ds^2 = \frac{L^2} {r^2} dr^2 + \frac{r^2}{L^2} \eta_{\mu \nu} dx^\mu
dx^\mu \, .
\end{gather}
The boundary is located at $r \rightarrow \infty$. 
The embedding of the Poincar\'e patch into global AdS is shown in
figure \ref{adsfig}.
A further choice of coordinates is obtained by introducing the coordinate $z \equiv L^2/r$, for
which the Poincar\'e metric \eqref{poin} becomes
\begin{gather}
ds^2 = \frac{L^2}{z^2} \left( dz^2 + \eta_{\mu \nu} dx^\mu dx^\nu
\right) \, .
\end{gather}
In this case, the boundary is located at $z \rightarrow 0$. 
There is a coordinate singularity at the boundary, but the space remains
regular there since the curvature remains finite. 

Note that the Ricci scalar and cosmological constant for Anti-de
Sitter space are both negative,
\begin{gather}
R = - \frac{d(d+1)}{L^2} \, , \qquad \Lambda = - \frac{d(d-1)}{2L^2}
\, .
\end{gather}

\subsection{Fisher information metric}

In the preceding section we have collected all the necessary
ingredients for formulating the AdS/CFT correspondence. Before now
proceding to see how the AdS/CFT conjecture arises within string
theory, we take a step back and for further motivation consider the
concept of {\it Fisher information}. This concept from statistical
mechanics gives rise to a metric on the space of probability
distributions. This provides a link between geometry and quantum-mechanical
probability densities in particular. We will use this to show how a
Gaussian probability distribution, which arises naturally for a quantum field
theory in the large $N$ limit, leads to an Anti-de Sitter
metric \cite{Matsueda:2013saa,Malek:2015hea,Blau:2001gj}.
 Relating AdS/CFT to concepts from information theory is a new
research direction which currently triggers a wealth of developments,
as discussed also in further lectures at this TASI 2017 School, see
for instance \cite{Harlow:2018fse}. 

Consider a normalized probability density with  $x$ as stochastic
variable and $\vec{\theta}$ a set of external parameters 
 A probability distribution $p(x, \vec \theta)$ is normalized to one,
\begin{gather}
\int \! dx p (x, \vec \theta) =1 \, ,
\end{gather}
and the expectation value of an operator $\mathcal{O}$  is given by
\begin{gather}
\langle \mathcal{O} \rangle  = \int \!dx \,  p(x,\vec\theta)
\mathcal{O}(x) \, .
\end{gather}
We define the spectrum $\gamma(x, \vec{\theta})  = - \ln p(x,
\vec \theta)$, which allosw us to write the von Neumann entropy  as
\begin{align}
S (\vec \theta) & = - \int \! dx p(x, \vec \theta) \ln
                  p(x,\vec\theta)\nonumber\\& = \int \! dx p(x,\vec
  \theta) \gamma (x, \vec \theta) = \langle \gamma \rangle \, .
\end{align}
The Fisher metric is then defined as
\begin{align} \label{Fisher}
g_{\mu \nu} (\vec \theta) = \int dx p(x, \vec \theta) \frac{\pr \gamma
  (x,\vec \theta)} {\pr \theta^\mu} \frac{\pr \gamma (x,\vec \theta)
  }{\pr \theta^\nu} = \langle \pr_\mu \gamma \pr_\nu \gamma \rangle \, .
\end{align}
Now let us evaluate this general expression for a Gaussian
distribution $p_G (x,\vec \theta)$ 
 as relevant for a saddle point approximation arising for instance in
 the large $N$ limit, 
\begin{gather}
p_G(x,\vec{\theta}) = p(x, \bar x, \sigma) = \frac{1}{\sqrt{2 \pi}
  \sigma} \exp \left(- \frac{(x-\bar x)^2}{2 \sigma^2} \right) \, ,
\end{gather}
where $\bar x$  is the mean of the distribution and  $\sigma$ its
standard deviation. For the Gaussian, the Fisher metric
\eqref{Fisher} becomes
\begin{gather}
g_{\mu \nu} d\theta^\mu d\theta^\nu = \frac{1}{\sigma^2} ( d \bar x^2+ 2\, 
d\sigma^2) \, .
\end{gather}
Subject to a suitable rescaling of the coordinates, this is the metric
of Euclidean AdS$_2$, with the standard deviation $\sigma$ playing the role of
the radial coordinate. The higher-dimensional case is obtained from a
distribution with several variables $x_1$, $\dots$, $x_d$, for which
\begin{gather}
p_G (x_1, \dots, x_d, \bar x_1, \dots, \bar x_d, \sigma) = \frac{1}{\sqrt{2 \pi} \sigma^d } \exp
\left( - \frac{\sum\limits_{i=1}^d(x_i - \bar x_i)^2}{2
    \sigma^2}\right)  \, .
\end{gather}
The Fisher metric for a Gaussian distribution is thus an AdS metric!
This is an interesting fact, however, so far we do not know what
determines the dynamics of this metric - there is no action on the
gravity side. So the relation found is not an example of a
gauge/gravity duality. 

With this additional motivation, let us now return to our example
where the dynamics is determined explicitly on both sides of the
duality, i.e.~the AdS/CFT correspondence as proposed by Maldacena.

%%%% CITATION

\subsection{String theory origin of the AdS/CFT correspondence}
\label{origin}

Given the motivation presented in the preceding subsection, let us
return to the duality example where the dynamics of the gravitational theory
{\it is} determined, i.e.~the duality  proposed
by Maldacena in 1997 \cite{Maldacena:1997re}. Let us consider this
duality for D3-branes within string theory.  In full generality, the conjecture states that
$\N=4$ $SU(N)$ Super-Yang-Mills theory is dual to type IIB string
theory on AdS$_5$ $\times$  $S^5$ for all values of $N$ and
$\lambda$. While this is a very beautiful idea, performing actual
explicit calculations for testing this proposal requires to consider
particular low-energy  limits which we will discuss in detail. This is
due to the fact that quantum string theory on curved backgrounds has
not yet been formulated. This is also a reason why it is hard to
provide an actual proof for the AdS/CFT proposal.

\subsubsection{Motivating AdS/CFT from string theory}

As a particular limit, we consider weakly coupled string theory with
string coupling $g_s \ll 1$, keeping $L/{\sqrt{\alpha'}}$ fixed. The
leading order is the classical string theory with $g_s=0$, which means
to only tree-level string diagrams are taken into account. On the CFT
side, since $ g_\mathrm{YM}^2 = 2 \pi g_s $ this implies $g_{YM}^2 =
\lambda /N \rightarrow 0$. This in turn implies that $N \rightarrow
\infty$ since $\lambda = L^4/ (2 \alpha'^2)$ remains finite. We are
thus considering the 't Hooft limit. The duality conjectured in this
limit, where $\lambda$ is fixed but may be small, and the dual field
theory contains classical strings, is often referred to as the {\it
  strong form} of the AdS/CFT correspondence. There is also the {\it
  weak form} of AdS/CFT in which additionally, $\lambda$ is taken to
be very large such that the CFT involved becomes strongly coupled. In
this case, the strongly coupled CFT is mapped to a classical
gravity theory of pointlike particles, since $\alpha' = \ell_s^2$
(with $\ell$ the string length) 
becomes asymptotically small. The gravity theory involved is type IIB
supergravity in the example considered. Type IIB supergravity admits
D3-brane solutions. The possible limits of the AdS/CFT correspondence
are collected together in table \ref{tableAdS}. 

%%%%% forms of AdS/CFT - table

\begin{table}[h]
{\begin{tabular}{|l|l|l|} \hline
 & $\N=4$ SYM theory & IIB theory on $AdS_5$ $\times$ $S^5$ \\ \hline \hline 
Strongest form &   any $N$ and $\lambda$ & Quantum string theory, 
$g_s \neq 0$, $\alpha' \neq 0$ \\ \hline  
Strong form & $N \rightarrow \infty$, $\lambda$ fixed but arbitrary & Classical
string theory, $g_s \rightarrow 0$, $\alpha' \neq 0$ \\ \hline  
Weak form & $N \rightarrow \infty$, $\lambda$ large & Classical
supergravity, $g_s \rightarrow 0$, $ \alpha' \rightarrow 0$ \\ \hline
  \end{tabular}}
\caption{Different forms of the AdS/CFT correspondence}
\label{tableAdS}
\end{table}

Let us now consider D3-branes to motivate the weak form of the AdS/CFT correspondence.
These branes may be viewed from two different perspectives: The open
and the closed string perspective.  It is crucial for the
correspondence that in the low-energy limit where only massless
degrees of freedom contribute, open strings give rise to gauge
theories while closed strings give rise to gravity theories.

{\bf Open string perspective.} We begin with the open string perspective on D3-branes. For $g_sN \ll
1$, D-branes may be visualised as higher-dimensional charged objectd
on which open strings may end. The `D' stands for Dirichlet boundary
condition. Consider a stack of $N$ D3-branes embedded in 9+1 flat
spacetime dimensions. (Recall that in 9+1 dimensions, superstring
theory is anomaly free and thus consistent.) Neumann and Dirichlet
boundary conditions are imposed on the string modes according to table
\ref{ND}. 

%%%%table ND
\begin{table}[h]
\begin{center}
{\begin{tabular}{|l|cccccccccc|}\hline
		&0&1&2&3&4&5&6&7&8&9\\ \hline
    $N \, D3$&$\bullet$&$\bullet$&$\bullet$&$\bullet$&-&-&-&-&-&- \\ \hline
  \end{tabular}}
\caption{Embedding of $N$ coincident D3--branes in flat ten-dimensional spacetime.}
\label{ND}
\end{center}
\end{table}
 
For $N$ coincident D3-branes, the open strings are described by a {\it
  Dirac-Born-Infeld (DBI) action} with gauge group $U(N)$,  with integration over the 3+1-dimensional
worldvolume of the branes. In flat ten-dimensional space, the  DBI action is given by
\begin{align}
S_\mathrm{DBI}  = & - T_3 \, \mathrm{tr} \, \int \! d^4x e^{-\varphi } \sqrt{
  - \mathrm{det} ( P[g] + 2 \pi \alpha' F)} \;\nonumber\\  &  + \, \mathrm{fermionic
  \; partners}  \, , 
\end{align}
where $T_3 \equiv 2/( (2\pi)^3 \alpha '^2 g_s)$ is the brane
tension, $\varphi$ is the dilaton, and $P[g]$ is the pullback of the
metric to the worldvolume of the branes. $F$ is the field strength
tensor of the gauge field associated to the brane charge.
We now consider low-energy excitations with $E
\ll \alpha'{}^{-1/2}$, such that only massless excitations are taken
into account. In this limit, the DBI action 
reduces to
\begin{align} 
S_\mathrm{DBI} &= - \frac{1}{2 \pi g_s} \; \mathrm{tr} \, \int\! d^4x
\, \left(  \frac{1}{2} F_{\mu \nu} F^{\mu \nu} +  \sum\limits_{i=1}^6 \pr^\mu
  \phi^i \pr_\mu \phi^i -  \pi g_s \sum\limits_{i,j=1}^{6}
              [   \phi^i, \phi^j]^2\right) \nonumber\\ & \hspace{5mm} 
\, + \, \mathrm{fermions} \, + \, {\cal O} (\alpha') \, , \label{DBIlow}
\end{align}
where the six scalars $\phi^i = \phi^{iA} T^A$ in the adjoint
representation of $U(N)$ arise from the pull-back of the metric to
the world-volume of the $N$ D3-branes. They are given by $X^{i+3} = 2
\pi \alpha' \phi^i$ with the $X^{i+3}$ the coordinates in the directions
perpendicular to the brane. 

The total action for the D3-branes is
\begin{gather} \label{SD3}
S_{D3} = S_\mathrm{DBI} + S_\mathrm{closed} + S_\mathrm{int} \, ,
\end{gather}
where $S_\mathrm{closed}$ describes the closed string excitations in
the ten-dimensional space and $S_\mathrm{int}$ the interaction between
open and closed string modes.
In the low-energy  limit $\alpha'
\rightarrow 0$, the open strings decouple from any closed string
excitations in the 9+1-dimensional space: In \eqref{SD3}, $ S_\mathrm{closed}$
becomes a free theory of massless metric fluctuations, and $S_\mathrm{int} $
goes to zero. In this limit we are thus left with the low-energy modes
in the DBI action as given by \eqref{DBIlow}, plus free massless gravity
excitations about flat space.  The low-energy  modes described by the
DBI action coincide
with the field-theory action of $\N=4$ Super Yang-Mills theory as
given by \eqref{N4action}, 
\begin{gather}
\lim\limits_{\alpha' \rightarrow 0} S_\mathrm{DBI} = S_{{\cal{N}} = 4
  \, \mathrm{SYM}} \, , 
\end{gather}
subject to identifying $2 \pi g_s = g_{YM}^2$. We thus recover the
action of $\mathcal{N} = 4$ Super Yang-Mills theory in this limit. By
modding out the center of the gauge group, we may reduce the $U(N)$
gauge symmetry to $SU(N)$. Note that the limit taken is $\alpha'
\rightarrow 0$ while keeping $u = r/\alpha'$ fixed, with $r$ any
length scale. This is referred to as the {\it Maldacena limit}. 

{\bf Closed string perspective.} We now turn to the closed string perspective on D-branes. In the limit
$g_s N \gg 1$, the $N$ D3-branes may be viewed as massive extended
charged objects sourcing the fields of type IIB supergravity. Closed
strings will propagate in this background. The supergravity solution
of $N$ D3-branes preserving $SO(3,1) \times SO(6)$ symmetry in
9+1 dimensions is given by
\begin{align}
ds^2 & = H(r)^{-1/2} \eta_{\mu \nu} dx^\mu dx^\nu + H(r)^{1/2}
  \delta_{ij} dy^i dy^j \, ,  \label{D3ansatz} \\
& e^{\varphi(r)} = g_s \, , \qquad C_{(4)} = \left( 1 - H(r)^{-1}
  \right) dx^0 \wedge dx^1 \wedge dx^2 \wedge dx^3 + \dots \, , \nonumber
\end{align}
with $\mu \nu \in \{ 0,1,2,3\}$ and $i,j \in \{1,2, \dots, 6\}$. 
Here, $r^2 = y_1^2 + y_2^2 + \dots + y_6^2$ and the terms denoted by the dots $\dots$ in
the expression for the four-form $C_{(4)}$ ensure self-duality of
$F_{(5)} = dC_{(4)}$, i.e.~the five-form given by the exterior
derivative of $C_{(4)}$. Inserting the ansatz \eqref{D3ansatz} into
the Einstein equations of motion in 9+1 dimensions, we find
that $H(r)$ must be harmonic,
i.e.
\begin{gather}
\bigtriangleup H(r) = 0 \, , \; \mathrm{for} \; r \neq 0 \, ,
\end{gather}
with $\bigtriangleup$ the Laplace operator in six Euclidean dimensions.
The Laplace equation is solved by
\begin{gather}
H(r) = 1+ \left(\frac{L}{r} \right)^4 \, .
\end{gather}
We will determine $L$ below.

Similarly to the open string case considered before, we now
investigate low-energy limits within the closed string perspective.
First we note that asymptotically for $r \rightarrow \infty$, we have $H(r)
\rightarrow 1$, i.e.~asymptotically for large $r$ we recover flat
9+1-dimensional space. On the other hand, there is the {\it
  near-horizon limit} in which $ r \ll L$. Then, $H(r) \sim L^4/r^4$
and the D3-brane metric becomes
\begin{align}
 ds^2 & =  \frac{r^2}{L^2} \eta_{\mu \nu} dx^\mu dx^\nu \, + \,
       \frac{L^2}{r^2} \delta_{ij} dy^i dy^j  \, , \nonumber\\ & =
                                                                 \frac{L^2}{z^2}
                                                                 \left(
                                                                 \eta_{\mu
                                                                 \nu}
                                                                 dx^\mu
                                                                 dx^\nu
                                                                 + dz^2
                                                                 \right)
                                                                 + L^2
                                                                 d\Omega_5^2
                                                                 \, ,
\end{align}
where in the second line we define the new radial coordinate $z \equiv
L^2/r$ and introduced polar coordinates on the space spanned by the six
$y^i$ coordinates, $dy^i dy^i = dr^2 = r^2 d \Omega_5^2$ with $d \Omega_5^2$ the
angular element on $S^5$. We see that in the near-horizon limit, the
D3-brane metric becomes AdS$_5$ $\times$ $S^5$!

$L$, i.e.~the radius of both the AdS$_5$ and the $S^5$, 
 may be determined from string theory. For this we note that the flux of
$F_{(5)}$ through the $S^5$ has to be quantized. The sphere $S^5$
surrounds the six Euclidean dimensions perpendicular to the D3-branes
at infinity. The charge $Q$  of the D3-branes is determined by
\begin{gather}
Q = \frac{1}{16 \pi G_{10}} \int\limits_{S^5} {}^* F_{(5)} \, .
\end{gather}
The charge has to coincide with the number of D-branes,
i.e. $Q=N$. This implies the important relation
\begin{gather}
L^4 = 4 \pi g_s N \alpha'^2 \, ,
\end{gather}
since $16 \pi G_{10} = 2 \kappa_{10}^2 = (2\pi)^7 \alpha'^4 g_s^2$. 

For stating the correspondence, we note that asymptotically, 
we observe two kinds of closed strings: Those in flat space at $r
\rightarrow \infty$, and those
in the near-horizon region. Both kinds decouple in the low-energy
limit. For an observer at infinity, the energy of fluctuations in the
near-horizon region is redshifted, 
\begin{gather}
E_\infty \sim \frac{r}{L} E_r  \rightarrow 0 \, .
\end{gather}
 Recall that $\sqrt{\alpha'}$  is fixed, but $r \ll L$. This implies
 that for an observer at infinity, the energy of fluctuations in the
 near-horizon region is very small. We thus have two types of massless
 excitations: Massless modes in flat space at $r \rightarrow \infty$
 and the modes in the near-horizon region, which appear as massless
 too.

{\bf Combining open and closed string perspectives.} The AdS/CFT correspondence is now motivated by identifying the
massless modes in the open and closed string perspectives.
First we note that as discussed above, both in the open and closed string pictures
there are massless modes corresponding to free gravity in flat
9+1-dimensional space. Moreover, in the open string picture further
massless modes are given by the Lagrangian of 3+1-dimensional
$\mathcal{N} = 4 $ $SU(N)$ Super Yang-Mills theory. On the other hand,
in the closed string picture we have gravity in the near-horizon
region, which is given by IIB supergravity on AdS$_5$ $\times$
$S^5$. Identifying these second types of massless modes in the open
and closed string pictures gives rise to the AdS/CFT conjecture.

As a final remark in this section, we note that in the near-horizon
limit of the closed string picture, it is not possible to locate the
D3-branes. In particular, it is not correct to state that they sit at
$r=0$. Rather, the D3-brane is a solitonic solution to 10d
supergravity which extends over all values of $r$ and which gives rise
to AdS$_5$ $\times $ $S^5$ in the near-horizon limit.

\subsubsection{Holographic principle}

{ An important feature of the AdS/CFT correspondence is that it is
  based on the holographic principle \cite{tHooft:1993dmi,Susskind}. 
Within semiclassical considerations for quantum gravity,
  the holographic principle states that the information stored in a spatial
  volume $V_{d}$ is encoded in its boundary area $A_{d-1}$, measured in
  units of the Planck area $l_p^{d-1}$.  This is motivated by the fact that
 the Bekenstein bound  applies to systems in which there is at most one
  degree of freedom per Planck area. The Bekenstein bound states that
  the maximum amount of entropy stored in a volume is given by $S =
  A_{d-1}/ (4G_{d+1})$, with its surface $A_{d-1}$ measured in Planck units and $G_{d+1}$ the Newton
  constant of the $(d+1)$-dimensional volume theory, including time.
 This leads in particular to the famous result that the
  entropy of a black hole is proportional to the area of its
  Schwarzschild horizon. 
The name `holographic principle' asserts that this principle is similar to a hologram as known
from optics, in which the information contained in a volume is stored on a surface.
}

\subsubsection{Field-operator map}
 
\label{sec:fom}

The argument given in section \ref{origin} motivates the conjectured
duality between a quantum field theory and a gravity theory. The map
between these two theories may be refined to a one-to-one map between individual
operators, i.e. between gauge invariant operators in $\N=4$ $SU(N) $
Super Yang-Mills theory  and classical gravity fields in AdS$_5$
$\times$ $S^5$. Each pair is given by identifying entries transforming in the same
representation of the  superconformal group $SU(2,2|4). $ 
The most prominent example are the 1/2 BPS or chiral primary operators
in the $[0, \Delta, 0]$ representation of the algebra of $SU(4)$. Here, the  three
entries are the Dynkin labels, with $\Delta$ the conformal dimension
of the corresponding operator.\footnote{A review of the group theory
  concepts mentioned here may for instance be found in appendix B of
\cite{Ammon:2015wua}.}
 The corresponding gauge invariant field
theory operators are
\begin{gather} \label{ODelta}
\mathcal{O}_\Delta (x) = \mathrm{Str} \left( X^{(i_1} (x) X^{i_2} (x) \dots
X^{i_\Delta)} (x) \right) = C^\Delta_{i_1 \dots i_\Delta} \mathrm{tr} \left(X^{(i_1} (x) X^{i_2} (x) \dots
X^{i_\Delta)} (x) \right) \, ,
\end{gather}
with the elementary real scalar fields $X^i$ as in
\eqref{N4action}. $\mathrm{Str}$ denotes the symmetrized trace over
the indices $(a,b)$ of the $SU(N)$ representation matrices
$T^A{}_a{}^b$. The symmetrization involves the totally symmetric $SU(4)$
rank $\Delta$ tensor representation $C^\Delta_{i_1 \dots
  i_\Delta}$. An important property of the 1/2 BPS  operators is that
their two- and three-point functions in $\N=4$ Super Yang-Mills theory
are not renormalized and thus independent of the 't Hooft coupling
$\lambda$. The perturbative small $\lambda$ results for these two- and
three-point functions may then directly be compared to their
counterparts calculated from the gravity side, which apply to
large $\lambda$.  However, since these correlation functions are
independent of $\lambda$, an exact matching of the field theory and
gravity results is expected and was indeed obtained in explicit
computations \cite{Freedman:1998tz,Lee:1998bxa}. This provides a non-trivial test of the
AdS/CFT proposal.

To obtain the corresponding fields on the supergravity side of the
correspondence, a Kaluza-Klein reduction is performed on $S^5$,
i.e.~the fields in ten dimensions are expanded in spherical harmonics on
$S^5$, which for a general scalar reads
\begin{gather}
\phi(x,z , \Omega_5) = \sum\limits_{l=0}^{\infty} \phi^l (x,z) Y^l
(\Omega_5)\, , \nonumber\\ \Box_{S^5} Y^l(\Omega_5)  = - \frac{1}{L^2} l (l+4)
Y^l (\Omega_5) \, .
\end{gather}
This Kaluza-Klein reduction of type IIB supergravity on $S^5$ was already performed in 1985 in \cite{Kim:1985ez}. 
From the Kaluza-Klein modes of the supergravity metric and five-form,
we may construct five-dimensional scalars $s^l(x,z)$ that are in the
same representation $[0, l, 0]$ as the field-theory operators
$\mathcal{O}_\Delta$ of dimension $\Delta$ defined in \eqref{ODelta}, subject to identifying $l = \Delta$. 
These scalars satisfy
\begin{gather}
\Box_{AdS_5} s^l (z,x)  = - \frac{1}{L^2} l (l-4)
s^l (x,z)\, .
\end{gather}
Asymptotically, near the AdS boundary at $z \rightarrow 0$, the
solutions to this equation satisfy
\begin{gather} \label{sl}
s^l (z,x) \sim s^l_{(0)} z^{4-l} + \langle \mathcal{O}_l \rangle
z^l + \, \mathrm{subleading \; terms} \, ,
\end{gather}
According to \cite{Gubser:1998bc}, the leading term $s^l_{(0)}$ may be identified with a
source for the 1/2 BPS operator $\mathcal{O}_l$, while the subleading term
involves the vacuum expectation value of this operator.\footnote{Note
  that in \eqref{sl}, $l$ is an index in representation space in
  $s^l$, however in $z^l$, $l$ is an exponent.}

For writing the AdS/CFT conjecture in terms of an equation, we add
sources for any gauge invariant composite operators to the CFT action,
\begin{gather}
S' = S - \int \! d^4x \phi_{(0)} (x) \mathcal{O}(x) \, .
\end{gather}
Wick rotating to Euclidean time, the generating functional for these
operators then reads
\begin{gather}
Z [ \phi_{(0)} ] = e^{- W[ \phi_{(0)}]} = \left\langle \exp \left(
    \int \!
  d^dx \, \phi_{(0)} (x) \mathcal{O} (x) \right)
\right\rangle_{\mathrm{CFT}} \, .
\end{gather}
The AdS/CFT conjecture may then be stated as
\begin{gather} \label{AdSCFT}
W[ \phi_{(0)} ] = S_{\mathrm{SUGRA}} [\phi ]  \Big|_{\lim\limits_{z
    \rightarrow 0} \left( \phi(x,z) z^{\Delta-4} \right)= \phi_{(0)}
    (x) } \, ,
\end{gather}
where $\Delta$ is the conformal dimension of the dual operator
$\mathcal{O}$. 
The boundary values of the supergravity fields are identified with the
sources of the dual field theory. Within AdS/CFT, the operator sources
of the CFT become dynamical classical fields propagating into the
AdS space in one dimension higher. Note also that AdS/CFT has elements
of a saddle point approximation since the CFT functional is given by a
classical action on the gravity side. This is expected in the large
$N$ limit which also amounts to a saddle point approximation.

\begin{figure}
\begin{center}
\includegraphics[scale=0.5]{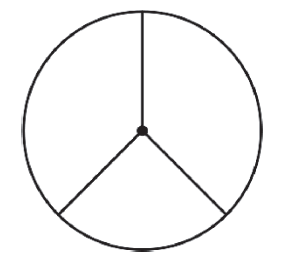}
\caption{Witten diagram for a three-point function.}
\label{3pt}
\end{center}
\end{figure}

From the proposal \eqref{AdSCFT} we may calculate connected Green's
functions in the CFT by taking functional derivatives with respect to
the sources on both sides of this equation. On the field theory side
we have
\begin{gather}
\langle \mathcal{O}_1 (x_1) \dots \mathcal{O}_n (x_n) \rangle = -
\frac{\delta^n W} {\delta\phi^1_{(0)} (x_1) \dots \delta\phi^n_{(0)}
  (x_n)} \Big|_{\phi^i_{(0)} = 0} \, .
\end{gather}
Using \eqref{AdSCFT} we may thus calculate CFT correlation functions from the propagation
of the source fields through AdS space. Since the gravity action is
classical, only tree diagrams contribute. The classical  propagators on
the gravity side are given by
the Green's functions of the operator $\Box_{\mathrm{AdS_5}}$, while
the vertices are obtained from higher order terms in the Kaluza-Klein
reduction of the ten-dimensional gravity fields on $S^5$. The
corresponding Feynman diagrams are referred to as Witten diagrams
\cite{Witten:1998qj} . These are usually drawn as a circle depicting the boundary
of AdS space, with the interior of the circle corresponding to the AdS
bulk space. An example for a Witten diagram leading to a three-point
function is shown in figure \ref{3pt}. Here, each of the three lines in the
bulk of AdS corresponds to bulk-to-boundary propagator, i.e. to the
appropriate Green's function of $\Box_{\mathrm{AdS_5}}$ with one
endpoint at the boundary. For scalar operators, the bulk-to-boundary
propagator is given by
\begin{equation} \label{KD}
K_\Delta (z_0, \vec{z}; \vec{x} ) = \frac{\Gamma (\Delta)}{\pi^{d/2}
  \Gamma(\Delta- \frac{d}{2}) }\left( \frac{z_0}{z_0^2 + (\vec{z} -
      \vec{x})^2 }\right)^\Delta
\end{equation}
in Euclidean AdS space with five-dimensional coordinates $z \equiv
(z_0, \vec{z}) $ with $z_0$ the radial coordinate and $\vec{z}$ the
four coordinates parallel to the boundary. For the second
coordinate, $x_0=0$ since $x$ is located at the boundary. 
The index $\Delta$ corresponds to the dimension of the dual
scalar operator. Moreover, 
the vertex in the Witten diagram corresponds to a cubic coupling obtained from  the
Kaluza-Klein reduction of the type IIB supergravity action on
$S^5$. For four-point functions or even higher correlation functions,
there are contributions involving bulk-to-bulk propagators that link
two vertices in the bulk of AdS space. The calculation of two- and
three point functions of 1/2 BPS operators in $\N=4$ Super Yang-Mills
theory and in IIB supergravity on AdS$_5$ $\times$ $S^5$
provides an impressive test of the AdS/CFT conjecture: The
results for the three-point function in field theory and gravity coincide, subject to an
appropriate normalization using the expressions for the two-point function \cite{Freedman:1998tz,Lee:1998bxa}.

\subsection{Finite temperature}

Let us now consider how the AdS/CFT correspondence may be generalized
to quantum field theory at finite temperature. In fact, there is a
natural way to proceed, which is based on the following. In thermal
equilibrium, quantum field theories may be described in the imaginary
time formalism. This means that the ensemble average of an operator at
temperature $T$ is given by
\begin{gather}
\langle {\cal O} \rangle_\beta = \mathrm{tr} \left( \frac{\exp (-\beta
    H)}{Z} \cal{O} \right)\, , \quad Z = \mathrm{tr} \exp (- \beta H)
\, ,
\end{gather}
where $\beta = 1/(k_B T)$ and we set $k_B =1$. $H$ is the Hamiltonian
of the theory considered. Formally, $\beta$ corresponds to an
imaginary time, $t= i \tau$. An important point is that the
analyticity properties of thermal Green's functions require $\tau \in
[0, \beta]$. This implies that the imaginary time $\tau$ is
compactified on a circle. 

Let us consider the gravity dual thermodynamics of $\N=4$ Super
Yang-Mills theory on $I \! \! R^3$. We note that the compactification
of the time direction breaks supersymmetry, since antiperiodic
boundary conditions have to be imposed on the fermions present in the
field theory Lagrangian. 

The essential point for constructing the gravity dual is that on the
gravity side, the field theory described above is identified with the
thermodynamics of black D3-branes in Anti-de Sitter space. 
The solitonic solution for these branes is given by the metric
\begin{gather}
ds^2 = H(r)^{-1/2} \left( - f(r) dt^2 + d \vec{x}^2 \right) \, + \,
H(r)^{1/2} \left( \frac{dr^2}{f(r) } + r^2 d \Omega_5{} ^2\right) \, ,
\\
f(r) = 1 - \left( \frac{r_H}{r} \right)^4 \, , \qquad H(r) = 1 +
\frac{L^4}{r^4} \, ,
\end{gather}
The blackening factor $f(r)$ vanishes
at the Schwarzschild horizon $r_h$ of the black brane. The difference
between a black brane and a black hole is that  the black brane is
infinitely extended in the spatial $\vec{x}$ directions, which span $I
\!\! R^3$.  Setting $z = L^2/r$, Wick rotating to imaginary time and
taking the near-horizon limit as before, this gives
\begin{gather}
ds^2= \frac{L^2}{z^2} \left( \left(1 -\frac{z^4}{z_H^4}\right) d\tau^2
  + d\vec{x}^2 + \frac{1}{1 -\frac{z^4}{z_H^4} } dz^2 \right) + L^2 d
\Omega_5^2  \, ,
\end{gather}
with $z_H$ the Schwarzschild radius. As for a black hole, we note that
$g_{\tau \tau} \rightarrow 0$, $g_{zz} \rightarrow \infty$ for $z
\rightarrow z_H$. Let us now introduce a further variable
\begin{equation}
z = z_H \left( 1 - \frac{\rho^2}{L^2} \right) \, .
\end{equation}
Here, $\rho$ is a measure for the distance from the horizon at $z_H$,
outside the black hole. We expand about the horizon. To lowest order
in $\rho$, the $(\tau,z)$ contribution to the Euclidean metric becomes
\begin{gather}
ds^2 \simeq \frac{4 \rho^2}{ z_H^2} d \tau^2 + d\rho^2 \, .
\end{gather}
With $\phi \equiv 2 \tau/z_H$, this becomes $ds^2 = d\rho^2 + \rho^2 d
\phi^2$. For regularity at $\rho=0$, we have to impose that $\phi$ is
periodic with period $2\pi$, such that we have a plane rather than a
conical singularity. This implies that $\tau$ becomes periodic with
period $\Delta \tau = \pi z_H$.  From the field-theory side we know
that $\Delta \tau = \beta = 1/T$, which implies
\begin{equation}
z_H = \frac{1}{\pi T}. 
\end{equation}
Thus the field-theory temperature is identified with the Hawking
temperature of the black brane!

We may now compute the field-theory thermal entropy from the
Bekenstein-Hawking entropy of the black brane
\cite{Gubser:1996de}. In general, the Bekenstein-Hawking entropy
is given by the famous result
\begin{equation}
S_{\mathrm{BH}} = \frac{A_{d-1}}{4 G_{d+1}} \, ,
\end{equation}
where $A_{d-1}$ is the area of the black brane horizon and $G_{d+1}$ is the
Newton constant. For a black D3-brane, the horizon area is given by
\begin{align}
A_3 &= \int \! d^3x \, \sqrt{g_{3d} \Big|_{z = z_H}} \cdot \mathrm{Vol}
    (S^5) \, , \qquad g_{3d} = g_{11} g_{22} g_{33} = \frac{L^6}{z^6}
      \, ,
    \nonumber\\ & = \pi^6 L^8 T^3 \mathrm{Vol} ( \mathbbm{R}^3) \, ,
\end{align}
where we used the useful formulae Vol$(S^5) = \pi^3 L^5$,
 \begin{equation} G_5 = 
\frac{G_{10} }{ \mathrm{Vol (S^5)}} = \frac{\pi L^3} {2 N^2} \, ,
\end{equation}
$2 \kappa_{10} = 16 \pi G_{10} = (2 \pi)^7 \alpha'^4 g_s^2$ and $L^4 =
4 \pi g_s N \alpha'^2$. Combining all results, we find
\begin{gather}
S_{\mathrm{BH}} = \frac{\pi^2}{2} N^2 T^3  \mathrm{Vol} (
\mathbbm{R}^3) \, .
\end{gather}
This result, valid at strong coupling, differs just by its prefactor
from the free field theory result
\begin{gather}
S_{\mathrm{free}} = \frac{2\pi^2}{3} N^2 T^3  \mathrm{Vol} (
\mathbbm{R}^3) \, .
\end{gather}
We note that the result at strong coupling is smaller by a factor of $3/4$.

\section{Holographic Kondo model}

\setcounter{equation}{0}

As an example of how to generalize the original example of the AdS/CFT
correspondence to more general cases, we will now study  how to obtain
a gravity dual of the well-known Kondo model of condensed matter
physics.

The original Kondo model \cite{Kondo} describes the interaction of a free electron
gas with a localized magnetic spin impurity. A crucial feature is that at low
energies, the impurity is screened by the electrons. 
The Kondo model is in agreement with experiments involving metals with
magnetic impurities, as it correctly
predicts a logarithmic rise of the resistivity as the temperature
approaches zero. 

The significance of the Kondo model goes far beyond its origin as a
model for metals with magnetic impurities. In particular, it played a
crucial role in the devopment of the renormalization group (RG). The
impurity coupling in the Kondo model has a negative beta function and
perturbation theory breaks down at low energies, a
property it shares with quantum chromodynamics (QCD). In some respects
the Kondo model may thus be viewed as a toy model for QCD. Moreover,
the Kondo model corresponds to a boundary RG flow connecting two RG
fixed points. These correspond to a UV and a IR CFT, respectively. CFT
techniques have proved very useful in studying the Kondo model, as
reviewed in
\cite{Affleck:1995ge}.
 Moreover, the Kondo model has a large $N$ limit in
which it may be exactly solved using the Bethe ansatz
\cite{Andrei,Wiegmann,Andrei:1982cr}. 

The holographic Kondo model we will introduce below differs from the
original condensed matter model in that the ambient electrons are strongly coupled among
themselves even before the interaction with the magnetic impurity isturned on. Moreover, the impurity is an $SU(N)$ spin with $N
\rightarrow \infty.$ The ambient degrees of freedom are dual to a
gravity theory in an AdS$_3$ geometry at finite temperature. The
impurity degrees of freedom are dual to an AdS$_2$ subspace. 
 As we will see in detail below, the dual gravity
model corresponds to a holographic RG flow dual to a UV fixed point
perturbed by a marginally relevant operator, which flows to an IR
fixed point. In addition, in the IR a condensate forms, such that the
model has some similarity to a holographic superconductor
\cite{Hartnoll:2008vx}.  For this
model, we may calculate spectral functions and compare their shape
to what is expected for the original Kondo model. This may be relevant
for the physics of quantum dots. Including the
backreaction of the impurity geometry on the ambient geometry allows
to calculate the entanglement entropy. Quantum quenches of
the Kondo coupling may also be considered.

\subsection{Kondo model within field theory and condensed matter
  physics}

Let us begin by considering the original model of Kondo \cite{Kondo}, which
describes the interaction of a free electron gas with a $SU(2)$ spin
impurity. The electrons are also in the spin 1/2 representation of a
second $SU(2)$. Using field-theory language, the corresponding 
Hamiltonian may be written as
\begin{equation} \label{KondoHamiltonian} 
H = \frac{v_F}{2 \pi} i \psi^\dagger \pr_x \psi \, + \,
\frac{v_F}{2}  \lambda_K \delta(x)  \vec{J}  \cdot \vec{S} \,   .\end{equation}
Here, $v_F$ is the Fermi velocity, and $\vec{S}$ is the magnetic
impurity satisfying
\begin{equation} \label{S}
\left[ S^a, S^b \right] = i \epsilon^{abc}S^c \, ,
\end{equation}
which takes values in the internal $SU(2)$ spin space. The spin impurity
interacts with the electron current
\begin{equation}
J^a = \psi^\dagger \sigma^a \psi \, ,
\end{equation}
with $\sigma^a$ the Pauli matrices. The
Hamiltonian consists of a kinetic term for the electrons and an
interaction localized at the site of the impurity. Hence the
interaction term involves a delta distribution. 

\begin{figure}
\begin{center}
\includegraphics[scale=0.8]{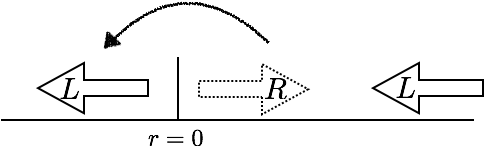}
\caption{Analytic continuation to negative values of $r$. The 
right-movers become left-movers travelling at negative values of $r$. }
\label{analytic}
\end{center}
\end{figure}

The Kondo model is simplified in the s-wave approximation, where the
problem becomes spherically symmetric. We thus introduce polar
coordinates $(r, \theta, \phi)$. The dependence on the two angles
becomes trivial and we are left with a 1+1-dimensional theory in the
space spanned by $(r,t)$. The radial coordinate $r$ runs from zero to
infinity. The impurity sits at the origin and provides
a boundary condition. The electrons separate into left- and right
movers. It is now convenient to analytically continue $r$ to negative
values. Then, the previous right-movers become left-movers travelling
at negative values of $r$, i.e.~$\psi_R(r) \rightarrow \psi_L(-r)$, as
shown in figure \ref{analytic}.

%%% Figure analytic continuation

\begin{figure}
\begin{center}
\includegraphics[scale=0.8]{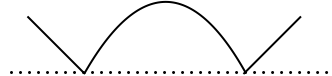}
\caption{One-loop Feynman graph contributing to the renormalization of
the Kondo coupling, with an electron (solid line) scattering off the
impurity (dashed line). }
\label{Feynmangraph}
\end{center}
\end{figure}

The Hamiltonian \eqref{KondoHamiltonian} was proposed and solved perturbatively
by Jun Kondo \cite{Kondo}. To first order in perturbation theory, the
quantum correction to the resistivity is
\begin{equation} \label{rho}
\rho(T) = \rho_0 \left[ \lambda_K + \nu \lambda_K^2 \ln \frac{D}{T} +
  \dots \right]^2 \, ,
\end{equation}
where $\nu$ is the density of states and $D$ a UV cut-off, for
instance the bandwidth. The corresponding Feynman graph is shown in
figure \ref{Feynmangraph}. 
 This correction explains the experimental
result for a logarithmic rise at low temperatures. From a theoretical
perspective, we note that perturbation theory breaks down at a
temperature scale
\begin{gather}
T_K =  D \, \exp \left( - \frac{1}{\nu \lambda_K}\right) \, ,
\end{gather}
which defines the {\it Kondo temperature} $T_K$. At this scale, the
first order perturbative correction  is of the same
order as the zeroth order term, which implies that perturbation theory
breaks down.

For the coupling itself, the first order perturbative correction gives
the beta function
\begin{gather} \label{betalambda}
\beta (\lambda_K) _{\mathrm{one-loop}}  = T \frac{d \lambda_K}{d T} = -
\nu \lambda_K^2 \, .
\end{gather}
So the beta function is negative. This is analogous to the gauge
beta function in QCD, which is also negative - a property associated
with asymptotic freedom in the UV. By analogy, we see that the Kondo
temperature $T_K$ plays a similar role as the scale
$\Lambda_\mathrm{QCD}$ in QCD, at which perturbation theory breaks down.

A resummation of  \eqref{betalambda} leads to the effective coupling
\begin{gather}
\lambda_{\mathrm{eff}} (T)= \frac{\lambda_K} {1 -\nu \lambda_K \ln
  (D/T)}  \, .
\end{gather}
$\lambda_{\mathrm{eff}} (T)$ diverges at $T\sim T_K = D \exp
(-1/(\nu \lambda_K))$.
In the IR for $T \rightarrow 0$, the theory has a strongly coupled
fixed point where the effective coupling vanishes. In fact, the
impurity is {\it screened}: The impurity spin forms a singlet with the
electron spin,
\begin{gather}
| \psi \rangle = \frac{1}{\sqrt{2}} \left( |  \Uparrow \downarrow \rangle - |
  \Downarrow \uparrow \rangle \right) \, .
\end{gather} 
This is reminiscent of the formation of meson bound
states in QCD. 

The theories at the UV and IR fixed points of the flow are described
by boundary conformal field theories (bCFT). Using the analytic
continuation described above, In the UV, the theory is free, and we
may impose the boundary condition $\psi_L(0) = \psi_R(0)$ for the
left- and right moving electrons introduced above. In the IR however,
due to the screening it costs energy to add a further electron to the
singlet at $r=0$. The probability for an electron to be at $r=0$ in
the ground state is zero. This observation is encoded in the
antisymmetric boundary condition $\psi_R(0) = - \psi_L(0)$. 
Within bCFT, the Kondo model was analyzed extensively by Affleck and
Ludwig \cite{Affleck:1990by}, 
 making non-trivial use of the appropriate representations of
the conformal and the spin Kac-Moody algebra. 

Both the UV and the non-trivial IR fixed point of the Kondo RG flow
may be described using CFT techniques. Essentially, the interaction
may be translated into a boundary condition at $r=0$. Let us sketch
this approach, considering a general $SU(N)$ spin group instead of the
$SU(2)$ considered above, as well as $k$ species (also called channels or flavours) of
electrons. In the UV, the boundary condition relating the left- and
right movers is just $\psi_L(0) = \psi_R (0)$.  In the IR, a bound
state involving the impurity spin forms, which is a singlet when
$N=k=2$. This implies that it costs energy to add another electron at
$r=0$, and the probability of finding another electron there is
zero. This is described by an antisymmetric wave function as provided
by the boundary condition $\psi_L(0) = - \psi_R(0)$. 

It may be shown \cite{Affleck:1995ge} that by introducing the currents
\begin{gather}
J_\mathrm{charge} = : \psi^{\dagger \, \alpha i} \psi_{\alpha i} : \,
, \qquad
J^a_\mathrm{spin} = : \psi^{\dagger \alpha i} T^a_\alpha{}^\beta
\psi_{\beta i}  : \, , \qquad J^A_\mathrm{channel} = :\psi^{\dagger \alpha i}
\tau^A_i{}^j \psi_{\alpha j} : \, ,
\end{gather}
where the colon denotes normal ordering, $T^a_\alpha{}^\beta$ are
$SU(N)$ generators and $\tau^A_i{}^j $ are $SU(k)$ generators, the
Kondo Hamiltonian may be written as
\begin{gather}
H = \frac{1}{2\pi(N+k)} J^a_\mathrm{spin} J^a_\mathrm{spin} +
\frac{1}{2\pi (k+N)} J^A_\mathrm{channel}  J^A_\mathrm{channel}  +
\frac{1}{4 \pi Nk} ( J_\mathrm{charge} )^2 + \lambda_K \delta(r) S^a
J^a_\mathrm{spin} \, .
\end{gather}
In the IR, by writing
\begin{gather}
\mathcal{J}^a_\mathrm{spin}= J^a_\mathrm{spin}  + \lambda_K \delta(r)
S^a \, ,
\end{gather}
the interaction term may be absorbed into a new current
$\mathcal{J}^a_\mathrm{spin}$. Written in terms of this new current,
the Hamiltonian again reduces to the Hamiltonian of the free theory
without interaction. The interaction is thus absorbed and replaced by
the non-trivial boundary condition discussed above.

At the conformal fixed points, the spin, channel and charge currents
may be expanded in a Laurent series,
\begin{gather}
J^a(z) = \sum\limits_{n \in \mathbbm{Z}} z^{-n-1} J^a_n \, .
\end{gather}
The mode expansions then satisfy Kac-Moody algebras,
\begin{gather}
[ J^a_n, J^b_m ] = i f^{abc} J^c_{n+m} + \frac{n}{2} k \delta^{ab}
  \delta_{m+n, 0} \, ,
\end{gather}
as shown here for the spin current with $SU(N)_k$ symmetry, where $k$
denotes the level of the Kac-Moody algebra. Similarly, for the
channels we have a $SU(k)_N$ symmetry. The total symmetry of the model
is $SU(N)_k \times SU(k)_N \times U(1)$. 
The representations of the two Kac-Moody algebras are fused in a
tensor product. The two different boundary conditions in the UV and in
the IR lead to different representations and thus operator spectra for
the total theory.

In the simplest example when the spin is $s = 1/2$ and there is only
one species of electrons, $k=1$, then in the IR a singlet forms. More
generally, a singlet is present when $2s= k$, which is referred to as
{\it critical} screening. When $k < 2s$, however, the impurity has
insufficient channels to screen the impurity completely, and there is
a residual spin of size $|s - k/2|$. This is referred to as {\it
  underscreening}. On the other hand, when $k> 2s$ there are too many
electron species for a critical screening of the spin, which leads to
non-Fermi liquid behaviour, a situation called {\it overscreening}.

\subsection{Large $N$ Kondo model}

%%%%%%%%%% More here %%%%%%%%%

As was found by condensed matter physicists in the eighties
\cite{ReadNewns,Coleman}, the Kondo
model simplifies considerably when the rank $N$ of the spin group is
taken to infinity.  In this limit, the interaction term $\vec{J} \cdot
\vec{S}$ reduces to a product $\mathcal{OO}^\dagger$ involving as
scalar operator $\mathcal{O}$, and the
screening corresponds to the condensation of $\mathcal{O}$. 
For comparison to gauge/gravity duality, it will be useful to consider
this large $N$ solution in which the Kondo screening appears as a condensation
process in $0+1$ dimensions. In the large $N$ limit, a phase
transition is possible in such low dimensions since long-range
fluctuations are suppressed. 
Moreover, there is an alternative large $N$ solution of the Kondo model using the Bethe
ansatz \cite{Andrei,Wiegmann}. 

%%%%%%%%%%%%%%%%%%

The large $N$ limit of the Kondo model involves $N \rightarrow
\infty$, $\lambda \rightarrow 0$ with $\lambda N$ fixed. The vector
large $N$ limit of the Kondo model provides information about the
spectrum, thermodynamics and transport properties everywhere along the
RG flow, even away from the
fixed points. $1/N$ corrections may be calculated.

We consider totally antisymmetric representations of $SU(N)$ given by
a Young tableau consisting of one column with $q$ boxes, $q<N$. We
write the spin in terms of Abrikosov pseudo-fermions $\chi$, which means that
we consider
\begin{equation} \label{Schi}
S^a = \chi^\dagger{}^i T^a{}_i{}^j \chi_j  \, , \qquad a = 1,2, \dots
, N^2 -1 \, ,
\end{equation}
with $\chi$ in the fundamental representation of $SU(N)$. A state in
the impurity Hilbert space is obtained by acting on the vacuum state
with $q$ of the $\chi^\dagger$. This gives rise to a totally
antisymmetric tensor product with rank $q$. Since \eqref{Schi} is
invariant under phase rotations of the $\chi$'s, there is an
additional new $U(1)$ symmetry. This implies that we need to impose a
constraint since considering the $\chi$'s instead of $S^a$ should not
introduce any new degrees of freedom. We impose
\begin{gather} \label{q}
\chi^\dagger \chi = q \, ,
\end{gather}
i.e.~the charge density of the Abrikosov fermions is given by the size
of the totally antisymmetric representation. Together with the
fermions $\psi$ of the Kondo model, we have a $SU(N)$ singlet operator
\begin{equation} \label{OO}
\mathcal{O} (t) \equiv \psi^\dagger \chi\, , \qquad \Delta_{\mathcal{O}} =
\frac{1}{2} \, .
\end{equation}
Now in the large $N$ limit, the Kondo interaction $\vec{J} \cdot
\vec{S}$ simplifies considerably as follows. We make use of the Fierz
identity \eqref{complete}.
%\begin{equation}
%\sum\limits_{a=1}^{N^2 -1} T^a{}_{ij} T^a{}_{kl} = \frac{1}{2} \left( \delta_{il} \delta_{jk} -
%  \frac{1}{N} \delta_{ij} \delta_{kl} \right)\, , \qquad i,j,k,l \in
%\{1, \dots N \} \, .
%\end{equation}
For the Kondo interaction  this implies
\begin{gather}
\lambda \delta(x) J^a S^a = \lambda \delta(x) (\psi^\dagger T^a \psi)
(\chi^\dagger T^a \chi) = \frac{1}{2} \lambda \delta(x) \left(
  \mathcal{O} \mathcal{O}^\dagger - \frac{q}{N} (\psi^\dagger \psi
  )\right) \, ,
\end{gather}
where for sufficiently small $q$ we may neglect the last term in the
limit $N \rightarrow \infty$. 

In the large $N$ limit, the Kondo coupling is thus the coupling of
a`double-trace' deformation $\mathcal{OO}^\dagger$, with two
separately gauge invariant operators $\mathcal{O}$ and
$\mathcal{O}^\dagger$. This is similar to double-trace operators where
two separately gauge-invariant operators are multiplied to each other.  
For operators involving fields in the adjoint representation, traces have
to be taken to generate gauge-invariant operators. Here however,
$\mathcal{O}$ is gauge invariant without trace, since both $\psi$ and
$\chi$ are in the fundamental of $SU(N)$. The operator
$\mathcal{OO}^\dagger$ is of engineering dimension one. As defect
operator, it is marginally relevant, i.e.~it is marginal at the
classical level, but quantum corrections make it relevant.

In the large $N$ limit, the solution of the field-theory saddle point
equations reveals a second order mean-field phase transition in which
$\mathcal{O} $ condenses: There is a critical temperature $T_c$
above which $\langle \mathcal{O} \rangle = 0 $ and below which 
$\langle \mathcal{O} \rangle \neq 0 $. The critical temperature $T_c$
is slightly smaller than the Kondo temperature $T_K$ and may be
calculated analytically. The condensate spontaneously breaks the $U(1)$ symmetry of
the $\chi$ fermions. $1/N$ corrections smoothen this transition to a
cross-over.

At large $N$, the Kondo model thus has similarity with 
superconductivity that is triggered by a marginally relevant
operator. This observation provides a guiding principle for
constructing a gauge/gravity dual  of the large $N$ Kondo model.

\subsection{Gravity dual of the Kondo model}

The motivation of establishing a gravity dual of the Kondo model is
twofold: On the one hand, this provides a new application of
gauge/gravity duality of relevance to condensed matter physics. On the other hand,
this provides a gravity dual of a well-understood field theory model
with an RG flow, which may provide new insights into the working
mechanisms of the duality. It is important to note that our
holographic Kondo model will have some features that are distinctly
different from the well-known field theory Kondo model described
above. Most importantly, the 1+1-dimensional electron gas will be
strongly coupled even before considering interactions with the
impurity. This has some resemblance with a Luttinger liquid coupled to
a spin impurity. Moreover, the $SU(N)$ spin symmetry will be gauged. 
The holographic Kondo model has provided insight into the entanglement
entropy of this system. Moreover, quenches of the Kondo coupling in
the holographic model provide a new geometric realization of the
formation of the Kondo screening cloud. It is conceivable that further
work will also lead to new insight into the Kondo lattice that
involves a lattice of magnetic impurities. The Kondo lattice is a
major unsolved problem within condensed matter physics. Preliminary
results in this direction that were obtained using holography 
may be found in \cite{Harrison:2011fs}. Further holographic studies of
holographic Kondo models include \cite{Benincasa:2012wu}. 
    
Here we aim at constructing a holographic Kondo model realizing similar features
to the ones of the large $N$ field theory Kondo model described in the
previous section, including a RG flow triggered by a doulbe-trace
operator \cite{Erdmenger:2013dpa}.  For this purpose, consider an appropriate configuration of D-branes which allows us
to realize the field theory operators needed.
The field theory
involves fermionic fields $\psi$ in 1+1 dimensions in the fundamental
representation of $SU(N)$, as well as Abrikosov fermion fields $\chi$
localized at the 0+1-dimensional defect. These transform in the
fundamental representation of $SU(N)$ as well. From these we will
construct the required operators. For the brane configuration we will use probe
branes, which means that a small number of coincident branes are
embedded into a D3-brane background, neglecting the backreaction on
the geometry. For a holographic Kondo model, a suitable choice of 
probe branes consists of D7- and D5-branes embedded as shown in table
\ref{table1}. 
%%% Table. Brane embeddings for the holographic Kondo model.
\begin{table}
\begin{center}
\begin{tabular}{|r|c|c|c|c|c|c|c|c|c|c|}
\hline
& 0 & 1 & 2 & 3 & 4 & 5 & 6 & 7 & 8 & 9 \\
\hline
$N$ D3    & X & X & X & X &   &   &   &   &   &   \\
\hline
$1$ D7 & X & X &    &   & X & X & X  & X   &  X &X   \\
\hline
$1$ D5 & X &  &    &   & X & X & X  & X   &  X &  \\
%\hline
%     & $z^+$ & $z^-$ & \multicolumn{2}{c|}{$x$, $\bar x$}
%     & \multicolumn{2}{c|}{$y$, $\bar y$} & $w^1$ & $w^2$ & $w^3$ & $w^4$
\hline
\end{tabular}
\caption{Brane configuration for a holographic Kondo model.}
\label{table1}
\end{center}
\end{table}

Fields in the fundamental representation are obtained
from strings stretching between the D3-, D5- and D7-branes. The
D7-brane probe extends in 1+1 dimensions of the worldvolume of the
D3-branes. As we discuss below, strings stretching between the D3- and D7-branes give rise
to chiral fermions, which we identify with the electrons of the Kondo
model. On the other hand, since the D5-brane only shares the time
direction with the D3-branes, the D3-D5 strings give rise to the 0+1
dimensional Abrikosov fermions.

We note that in a in absence of the D5-branes, the D3/D7-brane system
has eight ND directions, such that half of the original supersymmetry is
preserved. However, the D5/D7-system has only two ND directions, such
that supersymmetry is broken. This leads to the presence of a tachyon potential and a
condensation as required for the large $N$ Kondo model. The tachyon, a
complex scalar field $\Phi$, is identified as the gravity dual of the
operator $\mathcal{O} = \psi^\dagger \chi$.

As discussed in \cite{Gomis,Harvey}, the D7-brane gives rise to an
action
\begin{equation}
S_7 = \frac{1}{\pi} \int d^2 x \psi_L^\dagger ( i \pr_- - A_- ) \psi_L 
\end{equation}
of chiral fermions which are coupled to the $\N=4$ supersymmetric
gauge theory in 3+1 dimensions. $A_-$ is a restriction of a component
of the $\N=4$ Super Yang-Mills gauge field to the subspace of the
fermions.
 These fermions are in the fundamental
representation of the gauge group $SU(N)$. For simplicity, from now on
we drop the label $L$ for left-handed. The gauge field $A_-$  is a component
of the $\N=4$ theory gauge field on the 1+1-dimensional subspace
spanned by the D7-brane. We identify the $\psi_L$ with the electrons
of the Kondo model.

Similarly, for the Abrikosov fermions $\chi$ we obtain from the
D3/D5-brane system the action
\begin{gather} \label{d3d5}
S_5 = \int dt \chi^\dagger ( i \pr_t  - A_t - \Phi_9) \chi \, .
\end{gather}
Here, $\Phi_9$ is the adjoint scalar of $\N=4$ Super Yang-Mills theory
whose eigenvalues represent the positions of the D3-branes in the
$x^9$ direction. In \eqref{d3d5}, both $A_t$ and $\Phi_9$ are
restricted to the subspace of the $\chi$ fields. 
Note that unlike the original Kondo model, the $SU(N)$ spin 
symmetry is gauged in this approach. Also, the background $\N=4$
theory is strongly coupled in the gravity dual approach 
and provides strong interactions between the electrons.

Let us now turn to the gravity dual of this configuration. The $N$
D3-branes provide an AdS$_5$ $\times$ $S^5$ supergravity background as
before. The probe D7-brane wraps an AdS$_3$ $\times$ $S^5$ subspace
of this geometry, while the probe D5-branes wraps AdS$_2$ $\times$
$S^4$. The Dirac-Born-Infeld action for the D5-brane contains a gauge
field $a_\mu$ on the AdS$_2$ subspace spanned by $(t,r)$, with $t$ the
time coordinate and $r$ the radial coordinate in the AdS geometry. The
$a_t$ component of this gauge field is dual to the charge density of
the Abrikosov fermions, $q = \chi^\dagger \chi$. The D7-brane action
contains a Chern-Simons term for a gauge field $A_\mu$ on AdS$_3$. 
As noted before, the D5-D7 strings lead to a complex scalar tachyon
field.

We may thus establish the holographic dictionary for the operators of
the field-theory large $N$ Kondo model given in table \ref{table2}.
The electron current in 1+1 dimensions is dual to
the Chern-Simons field in 2+1 dimensions. The Abrikosov fermion charge
density $q$ in 0+1 dimensions is dual to  the gauge field component
$a_t$ in 1+1 dimensions. Finally, the operator $\mathcal{O} =
\psi^\dagger \chi$ in 0+1 dimensions is dual to the complex scalar
field $\Phi$ in 1+1 dimensions.

\begin{table}
\begin{center}
\begin{tabular}{|l|c|l|}
\hline
{Operator} & & {Gravity field} \\ \hline
Electron current $J^\mu  = \bar \psi \gamma^\mu \psi$ &$\Leftrightarrow$ & Chern-Simons gauge field $A$
in $AdS_3$ \\  \hline Charge density $q =
\chi^\dagger \chi$   & $\Leftrightarrow$ & 2d gauge field $a$ in $AdS_2$
\\ \hline
Operator ${\cal O} =
\psi^\dagger \chi$  & $\Leftrightarrow$ & 2d complex scalar $\Phi$ in
                                          AdS$_2$  \\
\hline
\end{tabular}
\caption{Field-operator map for the holographic Kondo model.}
\label{table2}
\end{center}
\end{table}

The brane picture has allowed us to neatly establish the required
holographic dictionary. Unfortunately, it is extremely challenging to
derive the full action describing the brane construction given. In
particular, the exact form of the tachyon potential is not known. 

For making progress towards describing a variant of the Kondo model
holographically, we thus turn to a simplified model consisting of a
Chern-Simons field in AdS$_3$ coupled to a Yang-Mills gauge field and
a complex scalar in AdS$_2$. This simplification still allows us to
use the holographic dictionary established above. The information we
lose though is about the full field content of the strongly coupled
field theory. On the other hand, this simplifield model allows for
explicit calculations of observables such as two-point functions and
the impurity entropy, as we discuss below. It is instructive to
compare the results of these calculations with features of the
field-theory large $N$ Kondo model, as we shall see. 

The simplified model we consider is
\begin{align}
S = &\frac{1}{8 \pi G_N} \int dz dx dt \sqrt{-g}\,  (R - 2 \Lambda) - \frac{N}{4
  \pi} \int\limits_{\mathrm{AdS}_3}  A \wedge {d} A \nonumber\\ & - N \int dx dt
                                                    \sqrt{-g} \left(
                                                    \frac{1}{4}
                                                    \mathrm{tr} f^{mn}
                                                    f_{mn}  + (D^m 
                                                    \Phi)^\dagger (
                                                    D_m \Phi) -
                                                    V(\Phi) 
                                                    \right) \, .
\end{align}
Here, $z$ is the radial AdS coordinate, $x$ is the spatial coordinate
along the boundary and $t$ is time. The defect sits at $x=0$. The
first term is the standard Einstein-Hilbert action with negative
cosmological constant $\Lambda$. The second term is a Chern-Simons
term involving the gauge field $A_\mu$ dual to the electron current
$J^\mu$. We take $A_\mu$ to be an Abelian gauge field, which implies
that we consider only one flavour of electrons, or - in condensed
matter vterms - only one channel.  $f_{mn}$ is the field strength
tensor of the gauge field $a_m$ with $m \in \{t,z \}$,  which we take
to be Abelian too. Its time component $a_t $ is dual to the charge
density $\psi^\star \psi$, which at the boundary takes the value $Q=
q/N$ with $q$ the dimension of the antisymmetric prepresentation of
the spin impurity.    $D_m$ is a covariant derivative given by $D_m =
\pr_m + i A_m \Phi - i a_m \Phi$. 
For the complex scalar, we assume its potential to take the simple
form
\begin{equation} \label{potK}
V (\Phi^\dagger \Phi) = M^2 \Phi^\dagger \Phi \, .
\end{equation}
We write the complex field as $\Phi = \phi \exp{i \delta}$ with $\phi
= |\Phi|$. We choose $M^2$ in such a way that $\Phi^\dagger \Phi$ is a
relevant operator in the UV limit. It becomes marginally relevant
when perturbing about the fixed point. Moreover, for the time being we consider the
matter fields as probes, such that they do not influence the
background geometry. For this background geometry we take the solution
to the gravity equations of motion which corresponds to the AdS BTZ
black hole, i.e.
\begin{gather} ds^2_{\mathrm{BTZ}} = \frac{1}{z} \left( \frac{1}{h(z)} dz^2 - h(z)
  dt^2 \right)\, , \nonumber\\ h(z) = 1 - \frac{z^2}{z_h^2} \, ,
\end{gather}
where we set the AdS radius to one, $L=1$, and
$z_h$ is related to the temperature by
\begin{gather}
T = \frac{1}{2 \pi z_h} \, .
\end{gather}
The non-trivial equations of motion for the matter fields are given by
\begin{gather}
\pr_z A_x  = 4 \pi \delta (x) \sqrt{g} g^{tt} a_t \phi^2 \, ,
\nonumber\\
\pr_z ( \sqrt{-g}  g^{zz} g^{tt} \pr_z a_t ) = 2 \sqrt{-g} g^{tt} a_t
\phi^2 \, , \nonumber\\
\pr_z ( \sqrt{-g} g^{zz} \pr_z \phi) = \sqrt{-g} g^{tt} a_t^2 \phi +
\sqrt{-g} M^2 \phi \, . \label{eomK}
\end{gather}
The three-dimensional gauge field $A_\mu$ is non-dynamical, but will
be responsible for a phase shift similar to the one observed in the
field-theory Kondo model.

Above the critical temperature $T_c$ where $\mathcal{O}$ dual to
the scalar field condenses,
we have $\phi = 0$. Then, asymptotically near the boundary, we have
$a_t(z) \sim \frac{Q}{z} + \mu$, where $\mu$ is a chemical potential
for the spurious $U(1)$ symmetry rotating the $\chi$'s. The charge density
is given by $ \chi^\dagger \chi  = NQ$, with $Q= q/N$. 

For generating the Kondo RG flow, we need to turn on the marginally
relevant `double-trace' operator ${\mathcal O} {\mathcal{O}^\dagger}$. We
  choose the mass $M$ in the potential such that the field $\phi(z)$
  is at the Breitenlohner-Freedman stability bound \cite{Breitenlohner:1982jf}. The
  asymptotic behaviour of $\phi(z)$ near the boundary is then
\begin{gather}
\phi(z) = \alpha z^{1/2} \ln (\Lambda z) - \beta z^{1/2} + {\cal O}
(z^{3/2} \ln (\Lambda z) \, .
\end{gather}
Following \cite{Witten:2001ua,Aharony:2001pa}, the gravity dual of a double-trace
perturbation is obtained by imposing a linear relation between
$\alpha$ and $\beta$,
\begin{gather}
\alpha = \kappa \beta \, .
\end{gather}
We choose $\alpha$ to correspond to a source for the operator ${\cal
  O}$, while $\beta$ is related to is vacuum expectation value.
The physical coupling $\phi(z)$ should be a RG invariant,
i.e.~invariant under changes of the cut-off $\Lambda$. This implies
\begin{gather}
\kappa = \frac{\kappa_0}{1 + \kappa_0 \ln (\Lambda_0/\Lambda)} \, .
\end{gather}
At finite temperature, we obtain the analogous result
\begin{gather}
\kappa_T = \frac{\kappa_0} {1 + \kappa_0 \ln (\Lambda z_h)}
\end{gather}
This expression for the coupling $\kappa_T$ diverges at the
temperature
\begin{gather}
T_K = \frac{1}{2 \pi} \Lambda e^{1/\kappa_0} \, ,
\end{gather}
where $T_K$ is the {\it Kondo temperature}.
A similar behaviour is observed in the condensed matter Kondo
models. Moreover, this behaviour bears some similarity to QCD, where
the coupling  becomes strong at a scale $\Lambda_{\mathrm{QCD}}$,
below which bound states provide the natural description of the degrees of freedom.  
Of course, in the holographic Kondo model there are {\it two} couplings, one between the electrons
themselves and secondly the Kondo coupling $\kappa_T$. While the first
is strong along the entire flow, $\kappa_T$ diverges at the Kondo
temperature and then becomes small again at lower temperatures, where
the condensate forms. 

\begin{figure}
\begin{center}
\includegraphics[scale=0.4]{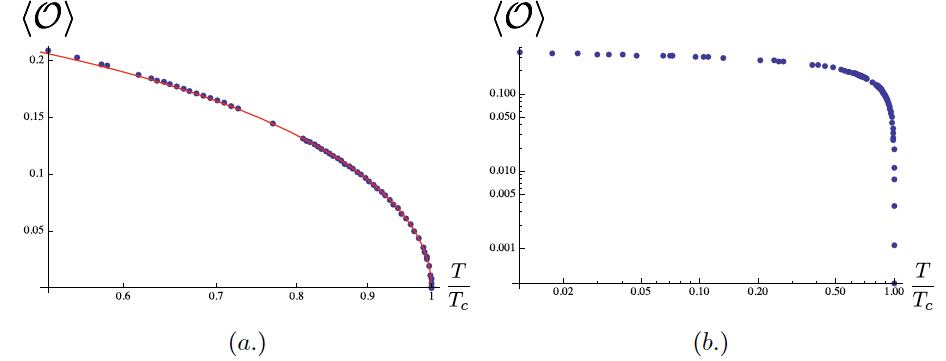}
\caption{Expectation value of the operator $\mathcal{O} =
  \psi^\dagger \chi $ as function of the temperature. Below $T_c$, a
  condensate forms. (a.) Close to the transition temperature,
  displaying that the phase transition is mean-field;  (b.)
  Log-log plot showing a larger temperature range. The VEV appears to
  approach a constant at low temperatures, however further
  stabilisation by a quartic potential contribution is expected to be
  required in the limit $T \rightarrow 0$.  Figures from
  \cite{Erdmenger:2013dpa}. }
\label{phasetransition}
\end{center}
\end{figure}

For determining the physical properties of the model considered, we
have to resort to numerics to solve the equations of motion
\eqref{eomK}. We find a mean-field phase transition as expected for a
large-$N$ theory, as shown in figure \ref{phasetransition}. In the
screened phase, a condensate of the operator ${\cal O} = \psi^\dagger \chi$
forms. We note that for very small temperatures, the numerical
solution of the equations of motion becomes extremely time-consuming
and thus our results are less accurate in this regime. We expect that
in the limit $T \rightarrow 0$, to obtain a stable constant solution
for $\langle \mathcal{O} \rangle$  requires to add a quartic term to
the potential \eqref{potK}.

%%% figure condensate 

\begin{figure}
\begin{center}
\includegraphics[scale=0.45]{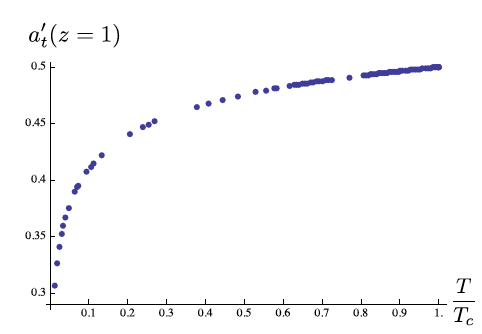}
\caption{Electric flux through the boundary of AdS$_2$ at the black hole horizon. This is a measure for
  the number of degrees of freedom. Its decrease at low temperatures
  indicates that the impurity is screened. For $T/T_c \lesssim 0.2$, the
  decrease is only logarithmic. The radial variable is normalized such
  that $z=1$ at the horizon. Figure from
  \cite{Erdmenger:2013dpa}.  }
\label{screening}
\end{center}
\end{figure}

Our holographic model allows for a geometrical description of the
screening mechanism in the dual strongly-coupled field theory. For
this we consider the electric flux $\mathcal{F}$ of the AdS$_2$ gauge field
$a_t(z)$. At the boundary of the holographic space, this flux encodes
information about the impurity spin representation,
\begin{gather}
\lim\limits_{z \rightarrow 0}  {\cal F} = \lim_{z \rightarrow 0} \sqrt{-g} f^{zt} = a'_t(z)|_{z \rightarrow 0}
= Q \, ,
\end{gather}
with $Q = q/N$ and $q$ as in \eqref{q}. 
When $\phi=0$, this flux is a constant and takes the same value at the
black hole horizon. However for $T < T_c$, the non-trivial profile 
$\phi(z)$ draws electric charge away from $a_t(z)$, reducing the
electric flux at the horizon. This implies that the effective number
of impurity degrees of freedom is reduced, which corresponds to
screening. This is shown in figure \ref{screening} which shows the
flux ${\cal F }_{z \rightarrow z_h}$ at the horizon as a function of
temperature. The numerical solution of the equations of motion yields
a decreasing flux when the temperature is decreased. 

The temperature dependence of the resistivity may be obtained by an
analysis of the leading irrelevant operator at the IR fixed point,
i.e.~by perturbing about the IR fixed point by this operator. This
gives $\rho(T) \propto T^\gamma$ with $\gamma \in \mathbbm{R}$ a real
number. A similar behaviour occurs also in Luttinger liquids
\cite{Luttinger}. The model thus does not reproduce the logarithmic
rise of the resistivity with decreasing temperature observed in the
original Kondo model. This behaviour is expected since the model is at
large $N$ and the ambient electrons are strongly coupled.

Let us emphasize again the differences between the holographic Kondo
model considered here and the large $N$ Kondo model of condensed
matter physics: Here, the electrons are strongly coupled among
themselves even before coupling them to the spin defect. The system
thus has two couplings: the electron-electron coupling which is always
large, and the Kondo coupling to the defect that triggers the RG flow.
Moreover, we point out that in our model, the $SU(N)$ symmetry is
gauged, while it is a global symmetry in the condensed matter models.

To conclude, let us consider different applications of the holographic
Kondo model we introduced. These involve three aspects: the impurity
entropy, quantum quenches and correlation functions. 

\subsection{Applications of the holographic Kondo model}

\subsubsection{Entanglement entropy}

\begin{figure}
\begin{center}
\includegraphics[scale=0.45]{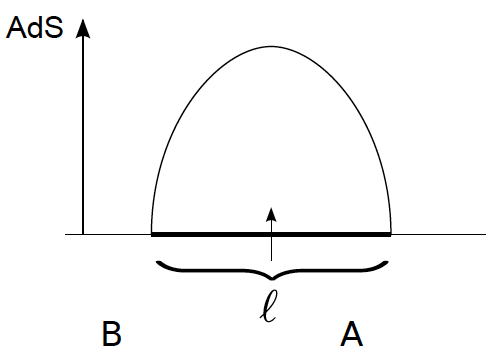}
\caption{The impurity entropy in the holographic Kondo model is
  obtained from the entanglement entropy. The entanglement area is a
  line of length $\ell$ in the dual field theory. The holographic
  minimal surface is a geodesic. For the impurity entropy, the
  entanglement entropy in absence of the defect is subtracted from the
one in presence of the defect.}
\label{entropy}
\end{center}
\end{figure}

The concept of holographic entanglement entropy introduced  by Ryu and
Takayanagi in 2006 has proved to be an important ingredient to the
holographic dictionary \cite{Ryu}, opening up new relations between gauge/gravity
duality and quantum information. In general, the entanglement entropy
is defined for two Hilbert spaces $\mathcal{H}_A$ and
$\mathcal{H}_B$. In the AdS/CFT correspondence, it is useful to
consider $A$ and $B$ to be two disjunct space regions in the
CFT. Defining the reduced density matrix to be
\begin{gather}
\rho_A = \tr_B \rho \, ,
\end{gather}
where $\rho$ is the density matrix of the entire space, the
entanglement entropy is given by its von Neumann entropy
\begin{gather}
S = - \tr_A \rho_A \ln \rho_A \, .
\end{gather}
The entanglement entropy bears resemblance with the black hole entropy
since it quantifies the lost information hidden in $B$. Ryu and
Takayanagi proposed the holographic dual of the entanglement entropy
to be
\begin{gather} \label{RT}
S = \frac{ \mathrm{Area} \gamma_A}{4 G_{d+1}}  \, ,
\end{gather}
where $G_{d+1}$ is the Newton constant of the dual gravity space and
$\gamma_A$ is the area of the minimal bulk surface whose boundary
coincides with the boundary of region A. For a field theory in 1+1
dimensions, the region A may be taken to be a line of length $\ell$,
and the bulk minimal surface $\gamma_A$  becomes a bulk geodesic joining the two
endpoints of this line, as shown for the holographic Kondo model in
figure \ref{entropy}. We note that for a 1+1-dimensional CFT
at finite temperature, with the BTZ black hole as gravity dual, it is
found both in the CFT \cite{Calabrese:2004eu} and on the gravity side
\cite{Ryu}  that the
entanglement entropy for a line of length $\ell$ is given by
\begin{gather} \label{SBH}
S_\mathrm{BH} (\ell) = \frac{c}{3} \ln \left( \frac{1}{\pi \epsilon T} \sinh (2 \pi
\ell T) \right) \, ,
\end{gather}
with $\epsilon$ a cut-off parameter.

\begin{figure}
\begin{center}
\includegraphics[scale=0.5]{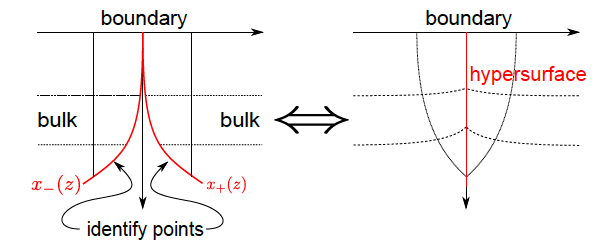}
\caption{Cutting and joining of two halves of the AdS BTZ geometry
  subject to the Israel junction at the defect.  Figure by Mario Flory. }
\label{ident}
\end{center}
\end{figure}

\begin{figure}[H]
\begin{center}
\includegraphics[scale=0.5]{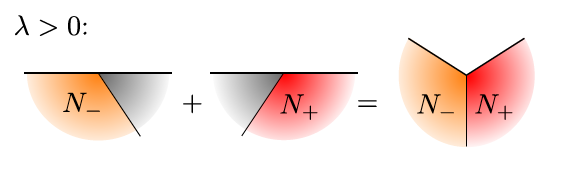}
\caption{Geometry in a vicinity of the backreacting defect brane at positive
  brane tension.  The horizontal black line corresponds to the
  boundary of the deformed AdS space, as in figure \ref{ident}. The volume is increased in a given region around the
  defect as compared to the case when the brane tension vanishes. This
will lead to a longer geodesic for a given entanglement interval and
thus to a non-zero positive impurity entropy. Figure by Mario Flory. }
\label{junction}
\end{center}
\end{figure}

For the Kondo model, a useful quantity to consider is the {\it impurity
  entropy} which is given by the difference of the entanglement
entropies in presence and in absence of the magnetic impurity,
\begin{gather} \label{Simp}
S_\mathrm{imp} = S_\mathrm{ impurity \; present} - S_\mathrm{impurity
  \; absent} \, . 
\end{gather}
In the previous sections, we considered the probe limit of the holographic Kondo model,
in which the fields on the AdS$_2$ defect do not backreact on the
AdS$_3$ geometry. However, including the backreaction is necessary in
order to calculate the effect of the defect on the Ryu-Takayanagi
surface. A simple model that achieves this
\cite{Erdmenger:2014xya,Erdmenger:2015spo} consists
of cutting the 2+1-dimensional geometry in two halves at the defect at
$x=0$ and joining these  back together subject to the {\it Israel
  junction condition} \cite{Israel:1966rt}
\begin{gather} \label{Israel}
  K_{\mu \nu}  - \gamma_{\mu \nu} K  = - \frac{\kappa_G}{2} T_{\mu \nu}
  \, ,
\end{gather}
This procedure is shown in figure \ref{ident}. We refer to the joining hypersurface as `brane'. 
In \eqref{Israel}, $\gamma$ and $K$ are the induced metric and extrinsic curvature
at the joining hypersurface extending in $(t,z)$ directions. $T_{\mu \nu}$
is the energy-momentum tensor for the matter fields $a$ and $\Phi$
at the defect, and $\kappa_G $ is the gravitational constant with
$\kappa_G^2 = 8 \pi G_N$. 
The matter fields $\Phi$ and $a$ lead to a non-zero tension on the brane, which
varies with the radial coordinate. The higher the tension on this
brane, the longer the geodesic joining the two endpoints of the
entangling interval will be, as shown in figure \ref{junction}.  A numerical solution of the Israel
junction condition reveals  that the brane tension decreases with
decreasing temperature, which leads to a shorter geodesic. This in
turn leads to a decrease of the impurity entropy \eqref{Simp}. This
decreases is expected and in agreement with the screening of the
impurity degrees of freedom.

In the holographic Kondo model, the brane is actually curved since the
brane tension depends on the radial coordinate. For large entangling
regions $\ell$, we may approximate the
impurity entropy to linear order by noting that the length decrease of
the Ryu-Takayanagi geodesic $\gamma_A$ translates into a decrease of the
entangling region $\ell$ itself. To linear order, this implies that
the entangling region is given by $\ell +D$ in the UV and by $\ell$ in
the IR, for $D \ll \ell$. Using \eqref{SBH} we may thus write for the
difference of the impurity between its UV and IR values
\begin{align}
\Delta S_\mathrm{imp} & = S_\mathrm{BH} (\ell +D) - S_\mathrm{BH}
                        (\ell) \nonumber\\
& \simeq D \cdot \pr_\ell S_\mathrm{BH} (\ell)  = \frac{2 \pi DT}{3}
  \coth (2 \pi \ell T)  \label{diff}
  \, .
\end{align}
It is a non-trivial result that subject to identifying the scale $D$
with the {\it Kondo correlation length} of condensed matter
physics, $D \propto \xi_K$, then the result agrees with previous
field-theory results for the Kondo impurity entropy
\cite{Laflorencie,Johannesson}. 

\subsubsection{Quantum quenches}

\begin{figure}
\begin{center}
\includegraphics[width=0.98\linewidth]{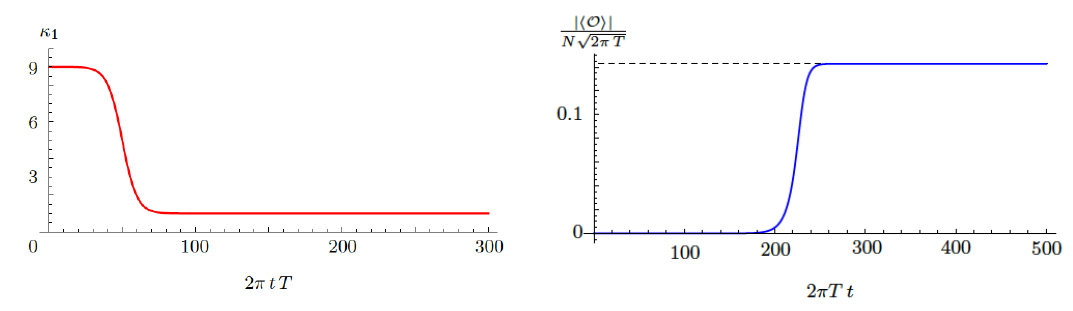}
\caption{Left: Quench of the `double-trace' Kondo coupling from the
  unscreened to the screened phase. Right: Reaction of the system to
  this quench: A condensate forms. There are no oscillations about the
  new equilibrium configuration. Figure from \cite{Erdmenger:2016msd}. }
\label{quench}
\end{center}
\end{figure}

A quantum quench corresponds to introducing a time dependence of the
Kondo coupling. On the gravity side, this implies that the equations
of motion become partial differential equations (PDEs), since both the
dependence on the AdS radial coordinate and on time are relevant.
Quenches of the holographic `double trace' Kondo coupling $\kappa_T$
were considered in \cite{Erdmenger:2016msd}. Figure \ref{quench} shows
a quench from the unscreened to the screened phase. The system reacts
to this quenchof the coupling by forming a condensate. There is a
certain time lapse before this happens. It is also noteworthy that the
reaction is overdamped, i.e. there are no oscillations around the new
equilibrium value. This behaviour follows from the structure of the
{\it quasinormal modes}, i.e.~the eigenmodes of the gravity
system. The leading eigenmode is purely imaginary in this system. This
is in agreement with the behaviour of the correlation functions
discussed in the next section.

\subsubsection{Correlation functions}

\begin{figure}
\begin{center}
\includegraphics[scale=0.75]{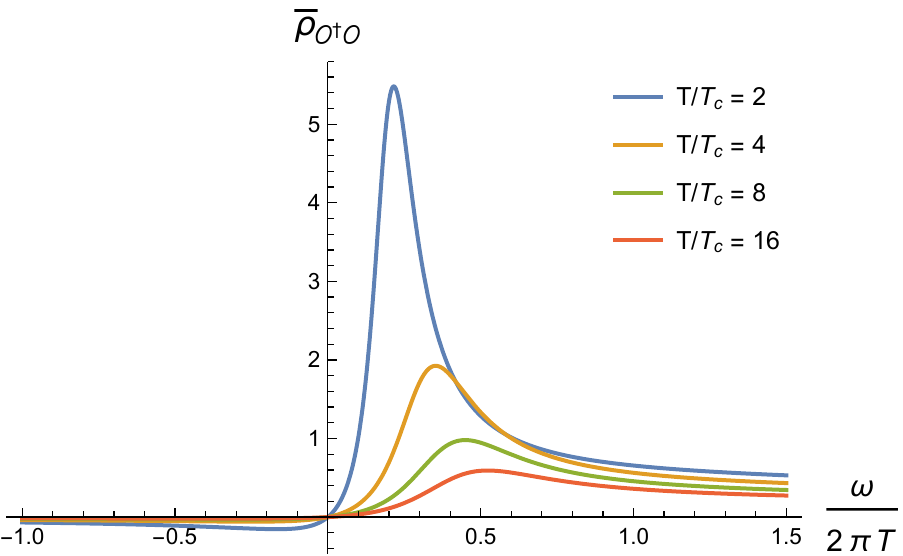}  \includegraphics[scale=0.75]{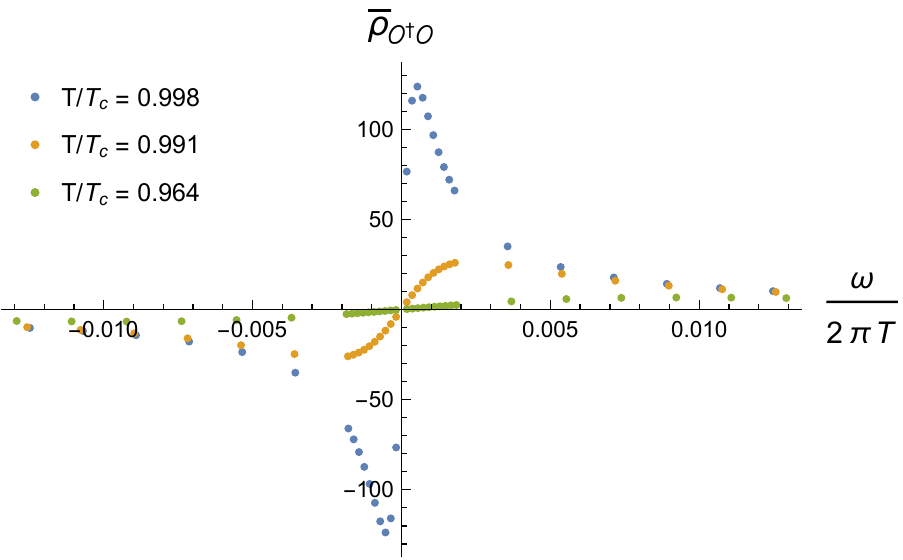} 
\caption{Spectral function $\rho(\omega)$ for the Kondo operator $\mathcal{O}$ at the
  defect, as function of the frequency $\omega$.  a) Left: In the unscreened phase above $T_c$. The spectral
  function corresponds to a Fano resonance with a spectral
  asymmetry. b) Right: In the screened phase below $T_c$. The spectral
  function is antisymmetric. The Green's functions' poles leading to
  the extrema in $\rho(\omega)$ are determined by the size of the
  condensate for $\mathcal{O}$. Figures from \cite{Erdmenger:2016vud}.}
\label{spec}
\end{center}
\end{figure}

AdS/CFT allows to calculate retarded Green's functions by adapting the
methods presented in section \ref{sec:fom} to Lorentzian signature
\cite{Son}. The required causal structure is obtained by imposing
infalling boundary conditions on the gravity field fluctuations at the
black hole horizon. Moreover, a careful regularization using the
methods of holographic regularization \cite{deHaro:2000vlm} is essential. This approach was used in
\cite{Erdmenger:2016vud,Erdmenger:2016jjg} to calculate spectral
functions for the Kondo operator $\mathcal{O}= \psi^\dagger \chi$ of
\eqref{OO}. Spectral functions are generally obtained from the
retarded Green's function by virtue of
\begin{gather}
\rho(\omega) = - 2 \, \mathrm{Im}  \, G_R(\omega) \, .
\end{gather}
The spectral function measures the number of degrees of freedom
present at a given energy. The results for the
holographic Kondo model obtained in
\cite{Erdmenger:2016vud,Erdmenger:2016jjg}  are shown in figure
\ref{spec}.

Above the critical temperature, these spectral functions show a 
{\it spectral asymmetry} related to a {\it Fano resonance}
\cite{Fano:1961zz}. In the holographic case, this asymmetry is characteristic
of the interaction between the ambient strongly coupled  CFT
and the localized impurity degrees of freedom. A similar
spectral asymmetry also appears in the condensed-matter large $N$
Kondo model (which involves free electrons) at vanishing temperature \cite{Parcollet:1997ysb}. In the screened
phase, the holographic spectral function displayed in figure
\ref{spec} is antisymmetric, consistent
with the relation
\begin{gather}
\omega_P \propto -i | \langle \mathcal{O} \rangle |^2 
\end{gather}
between the condensate and the leading pole $\omega_P$  in the retarded Green's
function. This relation is also satisfied by the condensed matter
large-$N$ Kondo model involving free electrons \cite{coleman_2015}. 

A similar spectral asymmetry also arises in the context of the Sachdev-Ye-Kitaev
(SYK) model that received a lot of attention recently
\cite{Sachdev:1992fk,Maldacena:2016hyu}. In fact, the original variant of
this model due to Sachdev and Ye \cite{Sachdev:1992fk} involves Weyl fermions, as
opposed to the Majorana fermions of the SYK model. This Sachdev-Ye may be obtained
from the Ising model by the same mechanism as discussed in \eqref{Schi} above,
i.e.~by writing the Ising spin in terms of a bilinear of auxiliary
fermions. In this case, the Ising model is given by
\begin{gather} \label{hising}
H_S = -  \frac{1}{\sqrt{N} } \sum\limits_{A<B} J_{A,B} {S}^{a A}
{S}^{a B}  \, , \qquad S^{a} = \psi^\dagger T^a \psi \, ,
\end{gather}
where the $A,B$ label the different sites of the Ising lattice, and
the index $a$ refers to spin space as in \eqref{Schi}. We see that
inserting the fermion bilinear expression for $S^a$ into the Ising
model will give rise to a four-fermion model. Indeed, as explained in 
\cite{Sachdev:1992fk,PhysRevB.63.134406},
reducing \eqref{hising} to a single-site model by  averaging over disorder, and
taking the large $N$ limit, gives rise to the Sachdev-Ye model
\begin{gather}
H_{SY} = \frac{1}{(2N)^{3/2} }\sum\limits_{i,j,k,l = 1}^{N}
  \overline{J_{ij,kl}} \, \chi^{\dagger  i} \chi^j \chi^{\dagger   k} \chi^l  -
  \mu \sum_i \chi^{\dagger  i} \chi^i \, ,
\end{gather}
where the second term involving the chemical potential $\mu$  is added to fix the representation $q$ of the
spin impurity.
As discussed in \cite{Sachdev:2015efa}, the Sachdev-Ye model also
displays a spectral asymmetry. This asymmetry is of an analogous form to the one
found above for the holographic Kondo model. In
\cite{Sachdev:2015efa}, it is shown that  the spectral asymmetry in the 
Sachdev-Ye model may be mapped to the entropy of a
black hole in AdS$_2$ space. A similar mechanism is expected to be at
work in the holographic Kondo model introduced above.

\section{Conclusion and outlook}

Though the concept of duality has existed for some time within
theoretical physics, the AdS/ CFT correspondence and its
generalizations to gauge/gravity duality are truly remarkable since
they relate a theory with gravity to a
quantum theory without gravity. This certainly added many new
viewpoints on fundamental questions such as the nature of quantum
gravity. On this basis, further significant progress is expected
within the next couple of years, one particular avenue being the quantum
physics of black holes and its relation to quantum information. This
provides a striking example of new developments in physics triggered
by joining different research areas that previously appeared as unrelated.
Equally striking is the new relation between fundamental and practical
questions provided by gauge/gravity duality, as it provides a new
approach for studying questions in strongly coupled quantum
systems. This has been applied to fields as diverse as elementary
particle, nuclear and condensed matter physics.

The holographic Kondo model demonstrates nicely how the original
concept of the AdS/CFT conjecture may be applied to more involved
configurations, in this case involving a marginally relevant
perturbation by a `double-trace' operator and a condensation process. 
 It also demonstrates that holographic models may be
linked to previous results, in this case the large $N$ Kondo model of
condensed matter physics. On the other hand, they also add new
features, in this case the coupling of the magnetic impurity to a
strongly coupled electron system, leading in particular to new
features in quantum quenches and in the spectral function.

The AdS/CFT correspondence and gauge/gravity duality are undoubtedly 
one of the most exciting developments in physics within the last twenty
years. As discussed, new avenues are opening up and are expected to
lead to further important discoveries in the future.

\bigskip

\section{Acknowledgements}

I am grateful to the organizers of TASI 2017, Mirjam Cvetic and Igor
Klebanov, for the opportunity to lecture at this lively and inspiring School, and to
the participants of TASI 2017 for pertinent questions. Moreover,
I would like to thank my co-author Martin Ammon for the joint writing
of the book \cite{Ammon:2015wua} that proved essential for preparing
these lectures. I am also grateful to my collaborators on the
holographic Kondo project for enjoyable and fruitful joint work: Mario Flory,
Carlos Hoyos, Ioannis Papadimitriou, Jonas Probst, Max Newrzella,
Andy O'Bannon and Jackson Wu.
Last but not least I would like to thank TASI 2017 participant Raimond
Abt for proofreading these lecture notes.

%\begin{thebibliography}{99}
% \bibitem{...} ....
%\end{thebibliography}

\bibliographystyle{utphys}
\bibliography{refs}{}

\end{document}